\documentclass[a4paper, fullpage]{article}

\usepackage{amsmath}
\usepackage{amssymb}
\usepackage{bm}
\usepackage{graphicx}
\usepackage{listings}
\usepackage{authblk}
\usepackage{color}
\usepackage{subcaption}
\usepackage[section]{placeins}

\newcommand{\vx}{\bm{x}}
\newcommand{\vt}{\bm{\theta}}
\newcommand{\vm}{\bm{m}}
\newcommand{\R}{\mathbb{R}}
\newcommand{\E}{\mathbb{E}}
\newcommand{\Var}{\mathbb{V}ar}
\renewcommand{\P}{\mathbb{P}}

\newcommand{\eq}[1]{Eq.~(\ref{eq:#1})}
\newcommand{\fig}[1]{Fig.~(\ref{fig:#1})}

\definecolor{codegreen}{rgb}{0,0.6,0}
\definecolor{codegray}{rgb}{0.5,0.5,0.5}
\definecolor{codepurple}{rgb}{0.58,0,0.82}
\definecolor{backcolour}{rgb}{0.99,0.99,0.97}
\lstdefinestyle{custom}{
  language=C++,
  literate={~}{$\sim$}{1},
  backgroundcolor=\color{backcolour},   
  commentstyle=\color{codegreen},
  otherkeywords = {real, vector, matrix, data, model, parameters, transformed},
  keywordstyle=\color{magenta},
  numberstyle=\tiny\color{codegray},
  stringstyle=\color{codepurple},
  emph={		normal, cauchy, inv_gamma, bernoulli_logit, gamma	},
  emphstyle=\color{codepurple},	basicstyle={\footnotesize,\ttfamily},
  breakatwhitespace=false,         
  breaklines=true,                 
  captionpos=t,                    
  keepspaces=true,                 
  numbers=left,                    
  numbersep=5pt,                  
  showspaces=false,                
  showstringspaces=false,
  showtabs=false,                  
  tabsize=2
}
\lstdefinestyle{customInline}{
  language=C++,
  literate={~}{$\sim$}{1},
  backgroundcolor=\color{backcolour},   
  commentstyle=\color{codegreen},
  otherkeywords = {real, vector, matrix, data, model, parameters, transformed},
  keywordstyle=\color{magenta},
  stringstyle=\color{codepurple},
  emph={		normal, cauchy, inv_gamma, bernoulli_logit, gamma	},
  emphstyle=\color{codepurple},	basicstyle={\ttfamily},
  breakatwhitespace=false,         
  keepspaces=true,                 
  numbers=left,                    
  numbersep=5pt,                  
  showspaces=false,                
  showstringspaces=false,
  showtabs=false,                  
  tabsize=2
}

\title{Reality-check for Econophysics: Likelihood-based fitting of
  physics-inspired market models to empirical data}

\author[1,2]{Nils Bertschinger}
\author[2]{Iurii Mozzhorin}
\author[3]{Sitabhra Sinha}

\affil[1]{Frankfurt Institute for Advanced Studies, Frankfurt am Main, Germany}
\affil[2]{Goethe University, Frankfurt am Main, Germany}
\affil[3]{Institute of Mathematical Sciences, Chennai, India}

\begin{document}

\maketitle

\begin{abstract}
  The statistical description and modeling of volatility plays a
  prominent role in econometrics, risk management and finance. GARCH
  and stochastic volatility models have been extensively studied and
  are routinely fitted to market data, albeit providing a
  phenomenological description only.

  In contrast, the field of econophysics starts from the premise that
  modern economies consist of a vast number of individual actors with
  heterogeneous expectations and incentives. In turn explaining
  observed market statistics as emerging from the collective dynamics
  of many actors following heterogeneous, yet simple, rather
  mechanistic rules.  While such models generate volatility dynamics
  qualitatively matching several stylized facts and thus illustrate
  the possible role of different mechanisms, such as chartist trading,
  herding behavior etc., rigorous and quantitative statistical fits
  are still mostly lacking.

  Here, we show how {\em Stan}, a modern probabilistic programming
  language for Bayesian modeling, can be used to fit several models
  from econophysics. In contrast to the method of moment matching,
  which is currently popular, our fits are purely likelihood
  based with many advantages, including systematic model comparison
  and principled generation of model predictions conditional on the
  observed price history.  In particular, we investigate models by
  Vikram \& Sinha and Franke \& Westerhoff, and provide a quantitative
  comparison with standard econometric models.
\end{abstract}

\section{Introduction}

Agent-based models of speculative behavior in financial markets are
nowadays able to replicate many stylized facts simultaneously. Thus,
providing an alternative to standard econometric models, offering
behavioral explanations of observed market statistics \cite{LuxABM,LeBaronABM}. Yet, estimation
of such models is still challenging and has mostly resorted to
simulation based methods striving to match selected moments of the
data \cite{FWfit,GhonghadzeGMM}. A notable exception is \cite{LuxSMC} which proposes and
investigates the use of sequential Monte-Carlo methods.

Here, we follow this line of research and utilize modern software
tools from machine learning and statistics to fit agent-based market
models. In particular, we employ {\em Stan} \cite{Stan}, a
probabilistic programming language for Bayesian modeling, to fit two
different agent-based models, namely from Vikram \& Sinha
\cite{VS} and Franke \& Westerhoff \cite{FWcontest}. We believe that Bayesian
estimation has many advantages as it allows to access parameter
uncertainties as well as to generate model predictions. Furthermore,
being based on the full model probability, including the likelihood,
different models can be systematically compared, e.g.~based on their
predictive likelihood on held-out data.


\section{Stan and Hamiltonian MCMC}
\subsection{Bayesian modeling}

In Bayesian modeling observed data\footnote{Vectors are denoted with
  bold symbols throughout the text.} $\vx = (x_1, \ldots, x_N)$ are
related to unobserved parameters/latent variables
$\vt = (\theta_1, \ldots, \theta_K)$ in terms of a joint probability
distribution with density $p(\vx, \vt)$. This density is usually
factorized as $p(\vx, \vt) = p(\vx | \vt) p(\vt)$, i.e.~into the
parameter likelihood and prior density. Inference then rests on Bayes
rule to obtain the density of the {\em posterior} distribution
\[ p(\vt | \vx) = \frac{p(\vx | \vt) p(\vt)}{p(\vx)} \] where the
normalization is given by
$p(\vx) = \int p(\vx | \vt) p(\vt) d\vt$. The posterior
summarizes the information obtained about the unobserved parameters
$\vt$ and combines the a-priori assessment of the modeler,
$p(\vt)$, with the information obtained from the data
$p(\vx | \vt)$. Thus, conceptually Bayesian estimation boils down to
a rather mechanical application of Bayes rule, once the full model
$p(\vx, \vt)$ is specified. Below, we will explain how this applies
to different agent-based models and discuss, in particular, the role
of prior choices in Bayesian modeling.

In practice, the normalization constant $p(\vx)$ of the posterior
density is often intractable, involving an integral over the parameter
space. Accordingly many approximation methods have been proposed which
either aim to approximate it with a tractable density or allow to draw
posterior samples from its unnormalized density. Hamiltonian
Monte-Carlo (HMC) sampling is an example of the latter approach. As
the well known Metropolis-Hastings algorithm it is an Markov chain
Monte-Carlo method, i.e.~producing a sequence of possibly correlated
samples. A comprehensive and readable introduction to HMC and its
properties can be found in \cite{HMC}. Here, a rather short overview
of the method should suffice.

\subsection{Markov Chain Monte-Carlo (MCMC)}

Consider a target density $p^*(\vt)$, e.g.~the posterior distribution
$p(\vt | \vx)$ from a Bayesian model. MCMC aims to construct a {\em
  transition density} $T(\vt' | \vt)$ which leaves the target
density invariant, i.e.
\[ p^*(\vt') = \int T(\vt' | \vt) p^*(\vt) d\vt \] Such a
transition density can then be utilized to draw a sequence of samples
$\vt_1, \vt_2, \ldots$ with
$p(\vt_2, \ldots | \vt_1) = \prod_{i = 1}^{\infty}
T(\vt_{i+1} | \vt_{i})$. The well known Metropolis-Hastings
algorithm uses two step in order to compute a suitable transition from
$\vt_i = \vt$:
\begin{enumerate}
\item Draw $\vt'$ from a proposal density\footnote{Choosing a suitable proposal density is a
    crucial step in the Metropolis-Hastings algorithm as it controls
    how effectively the resulting transitions can move across the
    sampling space.} $q(\vt' | \vt)$
\item Either retain the current sample, i.e.~$\vt_{i+1} = \vt$
  or transition to $\vt_{i+1} = \vt'$ with acceptance probability
  \[ a_{\vt' | \vt} = \min\left[ 1, \frac{p^*(\vt') q(\vt |
        \vt')}{p^*(\vt) q(\vt' | \vt)} \right] \]
\end{enumerate}
This so defined transition density not only leaves the target density
invariant, but, under suitable conditions, also ensures that the chain
converges to its unique invariant density starting from any initial
condition $\vt_1$ \cite{PRML}.

\subsection{Hamiltonian Monte-Carlo (HMC)}
While, in theory, the Metropolis-Hastings algorithm can produce
samples from the desired target density, especially in high-dimensions
it can suffer from slow convergence. This arises if the proposal
density is not well matched to the target density leading to either
many rejected steps or small random steps diffusing slowly across the
sampling space. HMC utilizes gradient information in order to generate
long sweeps of proposed states which are nevertheless accepted. To
this end, HMC samples from an augmented state space $(\vt, \vm)$ with
density
\begin{eqnarray*}
  p(\vt, \vm) & = & p(\vt) p(\vm | \vt) \\
              & = & e^{\log p(\vt) + \log p(\vm | \vt)} \\
              & = & e^{- H(\vt, \vm)}
\end{eqnarray*}
In analogy with physical systems, $\vm$ is considered as the momentum
of a particle at position $\vt$ and the {\em Hamiltonian}
$H(\vt, \vm) = - \log p(\vt) - \log p(\vm | \vt)$ decomposes into a sum of
potential and kinetic energy respectively. The Hamiltonian dynamics,
i.e.
\begin{eqnarray*}
  \dot{\vt} & = & \frac{\partial}{\partial \vm} H(\vt, \vm)
                     = - \frac{\partial}{\partial \vm} \log p(\vm | \vt) \\
  \dot{\vm} & = & - \frac{\partial}{\partial \vt} H(\vt, \vm)
                = \frac{\partial}{\partial \vt} \log p(\vt) + \frac{\partial}{\partial \vt} \log p(\vm | \vt)
\end{eqnarray*}
then preserves total energy/probability and thus leads to new states
$(\vt', \vm')$ which can always be accepted. In practice, the above
differential equation needs to be integrated numerically and care has
to be taken that numerical errors do not accumulate. Fortunately,
symplectic integrators can efficiently integrate Hamiltonian systems
as numerical errors cancel and simulated trajectories closely
approximate the theoretical dynamics. Nevertheless, as energy is only
approximately conserved along numerical trajectories, in practice, HMC
uses a Metropolis-Hastings step to either accept or reject the final
position of a trajectory. Furthermore, before each transition a new
momentum is sampled according to $p(\vm | \vt)$ which is commonly
taken as a Gaussian distribution independent of the current state
$\vt$, i.e.~$\vm \sim \mathcal{N}(\bm{0}, \bm{I})$. 

Especially in high-dimensional models, i.e.~with many parameters, the
use of gradient information to guide exploration is crucial to ensure
efficient sampling. Note that HMC is restricted to continuous
parameter spaces $\vt \in \R^K$, but could be combined with other
methods when discrete parameters are desired. Often, it is
advantageous to marginalize over discrete parameters as strong
correlations between them can severely hinder efficient
sampling\footnote{This especially applies to Gibbs sampling. Despite
  its popularity Gibbs sampling is severely hindered by correlations in
  the posterior which can render it utterly useless in
  high-dimensional problems.}. In theory, HMC is insensitive to strong
correlations between parameters $\vt$. In practice, the symplectic
integrator uses a finite step size to numerically solve the
Hamiltonian dynamics. In the case of posterior densities with high
curvature this can prevent the sampler to reach certain parts of the
state space. Furthermore, it makes HMC sensitive to the scale of
parameters as the step size would need to be adjusted
accordingly\footnote{In contrast, Gibbs sampling is severely effected by
  strong dependencies but insensitive to the scale of
  parameters.}. Fortunately, by reparameterizing the model it is often
possible to simplify the geometry of the posterior density. Furthermore,
a range of diagnostics, e.g.~based on the stability of the numerical
trajectory, have been developed \cite{HMC}.

\subsection{The Stan language}
The probabilistic programming language {\em Stan} \cite{Stan} allows
the user to describe the joint probability $p(\vx, \vt)$ of a model in
a high-level programming language. In turn, the program is then
compiled to C++ and several inference algorithms, including
Hamiltonian Monte-Carlo, are build-in. The required gradients are
computed via a C++ library for automatic differentiation
\cite{StanMath}, thus freeing the user from manually implementing and
debugging gradient calculations.

{\em Stan} comes with an extensive documentation which includes many example
models and useful tricks for efficiently implementing them
\cite{StanDoc}. A minimal {\em Stan} program consists of three blocks
\begin{enumerate}
\item a {\bf data} block which declares variables corresponding to
  observed quantities $\vx$,
\item a {\bf parameters} block which declares variables corresponding
  to unobserved parameters $\vt$
\item and a {\bf model} block which contains statements computing the
  log density of the model, i.e.~$\log p(\vx, \vt)$.
\end{enumerate}
For instance, the following {\em Stan} program estimates the mean of
Gaussian observations with a known standard deviation:
\begin{lstlisting}[style=custom]
  data {
    int<lower=0> N; // number of data points
    vector[N] x;    // observed data points
  }
  parameters {
    real mu; // unobserved mean
  }
  model {
    mu ~ normal(0, 10); // computes log prior density log p(mu)
    x ~ normal(mu, 1);  // computes log likelihood log p(x | mu)
    // total log density is log p(mu) + log p(x | mu)
  }
\end{lstlisting}
The $\sim$ statements in the model block relate a variable with a
density and are short-hand notation for the more basic statements
summing up log density contributions, e.g.
\lstinline[style=customInline]{target += normal_lpdf(x | mu, 1)}.

As shown in the example, all variables are typed and need to be
declared before use. A particularly convenient feature of {\em Stan} is that
variables can be given constraint types, e.g.~a standard deviation
parameter could be declared as \lstinline[style=customInline]{real<lower=0> sigma}.
Internally, the variable is then automatically transformed
to an unbounded space and the log density adjusted for the resulting
change of measure. Due to this method many different data types
including vectors, matrices but also constraint spaces such as
simplices or covariance matrices are readily supported.

{\em Stan} supports several inference algorithms, namely gradient descent
optimizers for maximum a-posteriori estimation, HMC sampling and
stochastic gradient variational Bayes. While HMC is the least
efficient of these algorithms it usually provides the closest
approximation to the true posterior distribution. Furthermore, during
warmup, also known as burn-in, {\em Stan} adapts several parameters of the
algorithm such that the algorithm appears essentially parameter-free to the
user. This is especially effective for the No U-Turn Sampler (NUTS)
which automatically adjust the length of simulated trajectories. In a
nutshell NUTS integrates the Hamiltonian dynamics until it starts
turning back towards itself which is locally decided based on the
gradient direction. Care needs to be taken to ensure that the
resulting transitions leave the target density invariant. To this end,
trajectories are expanded in a tree like fashion by successively
doubling their length forward and backward in time. The next state is
then sampled uniformly from the resulting overall trajectory.

For further details about the {\em Stan} programming language and the NUTS
algorithm we refer the interested reader to the {\em Stan} manual
\cite{StanDoc} and \cite{HMC} respectively. Next, we turn to
agent-based models for financial markets and show how these can be
expressed as statistical models.

\section{Market models}

Financial markets exhibit some remarkable and often surprisingly
stable statistical signatures, often referred to as {\em stylized
  facts} \cite{ContStyle,LuxABM}. Most notable and researched are the properties of asset
price returns exhibiting fat-tailed distributions and volatility
clustering. Especially volatility, i.e.~the standard deviation of
returns, has received much attention in the econophysics community for
its auto-correlation decaying as a power-law suggesting a long-memory
process. Accordingly many models to accurately model its statistics or
explain the origin of volatility correlations have been developed.

Agent-based models consider the statistical signatures of financial
markets as emergent properties, i.e arising from the collective
actions of many interacting traders. Thus, complementing standard
economic models which, presuming rational actors, are often unable to
explain the rapid changes in volatility with calm market phases interrupted
by highly volatile episodes. Shiller has coined the term excess
volatility hinting at these shortcomings \cite{Shiller}. In contrast,
agent-based models allow for bounded rational actors and can often
reproduce the stylized facts presuming chartist trading and/or herding
behavior \cite{LuxAgent}.

Here, we consider two models namely by Vikram \& Sinha \cite{VS} and
Franke \& Westerhoff \cite{FWcontest} in detail. In particular, we explain
how these models give rise to a latent state dynamics which can be
simulated and estimated with Bayesian methods.

\subsection{Model by Vikram \& Sinha (VS)}

The market in the VS model is populated by $N$ traders. At each
time step $t$ a trader $i$ either buys ($S_i(t) = 1$), sells
($S_i(t) = -1$) or stays inactive ($S(t) = 0$). The normalized net
demand from all traders is then given as
$M_t = \frac{1}{N} \sum_{i=1}^N S_i(t)$ and the price adjusts as
$p_{t+1} = \frac{1 + M_t}{1 - M_t} p_t$. An agents decision to
buy/sell or staying out depends on the perceived mispricing between
the current price $p_t$ and its running average
$p^*_t = \langle p_t \rangle_{\tau}$ which is considered as a proxy
for the fundamental price of the asset. The probability of an agent to
trade is then given by
\[ \P(|S_i(t)|) = \exp^{ - \mu \left| \frac{p_t - p^*_t}{p^*_t}
  \right| } \]
and a trading agent buys $S_i(t) = 1$ or sells $S_i(t) = -1$ at
random with equal probability.

In order to obtain a statistical model of volatility, in particular
with a continuous latent state as required for HMC sampling, we have
adapted the model as follows:
\begin{itemize}
\item For a large number of agents $N \to \infty$ the net demand $M_t$ converges to a Gaussian distribution with
  mean zero (as $\E[S_i(t)] = 0$) and variance
  $\frac{\P(|S_i(t)| = 1)}{\sqrt{N}}$
  \begin{itemize}
  \item Here, we have used that agents trading decisions $S_i(t)$ are
    independent and
    {
      \footnotesize
      \begin{eqnarray*}
        \E[S_i(t)] & = & \frac{1}{2} \P(|S_i(t)| = 1) \cdot 1 + \frac{1}{2} \P(|S_i(t)| = 1) \cdot (- 1) + (1 - \P(|S_i(t)| = 1) \cdot 0 = 0 \\
        \Var[S_i(t)] & = & \E[S_i(t)^2] \\
                   & = & \frac{1}{2} \P(|S_i(t)| = 1) \cdot 1^2 + \frac{1}{2} \P(|S_i(t)| = 1) \cdot (- 1)^2 + (1 - \P(|S_i(t)| = 1) \cdot 0^2  \\
                   & = & \P(|S_i(t)| = 1)
      \end{eqnarray*}
    }
  \item Next, considering the number of agents as unknown we introduce
    a scaling parameter $\sigma_{max}^2$ for the variance and model
    the demand as
    $M_t \sim \mathcal{N}(0, \sigma_{max}^2 \P(|S_i(t)| = 1))$.
  \end{itemize}
\item Finally, we approximate the log-return by linearizing the price
  impact\footnote{We have also fitted the exact model, i.e.~putting a
    normal distribution on the transformed returns
    $M_t = \frac{e^{r_{t+1}} - 1}{e^{r_{t+1}} + 1}$ without any
    noticeable difference.}
  \begin{eqnarray*}
    r_{t+1} & = & \log \frac{p_{t+1}}{p_t} \\
            & = & \log \frac{1 + M_t}{1 - M_t} \\
            & \approx & 2 M_t
  \end{eqnarray*}
  where we have used that $\log (1 + x) \approx x$ for $|x| \ll 1$.
\end{itemize}
Overall, we arrive at the following model dynamics\footnote{Simulating
  this approximate model shows that it produces similar price series
  with strong volatility clustering as the original model.}
\begin{eqnarray}
  \langle p_{t} \rangle_{\tau} & = & (1 - \tau) p_{t} + \tau \langle p_{t-1} \rangle_{\tau} \nonumber \\
  \P(|S(t)| = 1) & = & e^{- \mu \left| \log \frac{p_t}{\langle p_t
                       \rangle_{\tau}} \right| }
  \label{eq:VS} \\
  r_{t+1} & \sim & \mathcal{N}(0, \sigma_{max}^2 \cdot 2 \P(|S(t)| = 1)) \nonumber
\end{eqnarray}
Note that this is a state-space model with a continuous latent state driving the time varying volatility
$\sigma_{t+1} = \sqrt{\sigma_{max}^2 \cdot 2 \P(|S(t)| = 1)}$.  Indeed, the
famous GARCH(1, 1) (generalized auto-regressive conditional
heteroscedastic) model \cite{GARCH} is of a similar form
\begin{eqnarray}
  \sigma_{t+1}^2 & = & \mu + \alpha r_t^2 + \beta \sigma_t^2 \nonumber \\
  r_{t+1} & \sim & \mathcal{N}(0, \sigma_{t+1}^2) \label{eq:GARCH}
\end{eqnarray}
The main difference between the VS (in our formulation) and the GARCH
model is that the volatility is a function of past prices in the former
and past returns in the latter model. Furthermore, due to being
founded in an agent-based model all parameters of the VS model are readily
interpretable as the sensitivity $\mu$ of the agents to mispricing and
the weighting $\tau$ of the running price average. In contrast, parameters
in the GARCH model are motivated purely from statistical grounds and
cannot easily be related to agent behaviors.

From a Bayesian perspective, \eq{VS} and \eq{GARCH} correspond to the
likelihood $p(\vx | \vt)$, i.e.~the conditional probability of the
observed data given the model parameters. To complete the model
density $p(\vx, \vt)$ we need to specify a prior distribution on the
parameters. The choice of a prior distribution is often considered as
subjective (whereas the likelihood has an aura of
objectivism). Arguably, from the perspective of modeling the observed
data this distinction is of limited relevance. Instead, note that fixing
the prior implicitly fixes a distribution on the data space,
i.e.~obtained as $p(\vx) = \int p(\vx, \vt) d\vt$ by marginalizing over
the parameters. A model can be considered as misspecified when it
assigns very low probability to the actual observed data. In contrast,
a good model should be able to generate similar data with reasonable
probability. This viewpoint is in line with \cite{GelmanPrior} who
argue that the prior can only be understood in the context of the
likelihood. Indeed prior and likelihood act together in shaping the
model and expressing our expectation about plausible data.

\begin{figure}[ht]
  \centering
  \includegraphics[width=1\textwidth, trim=0 120 0 120]{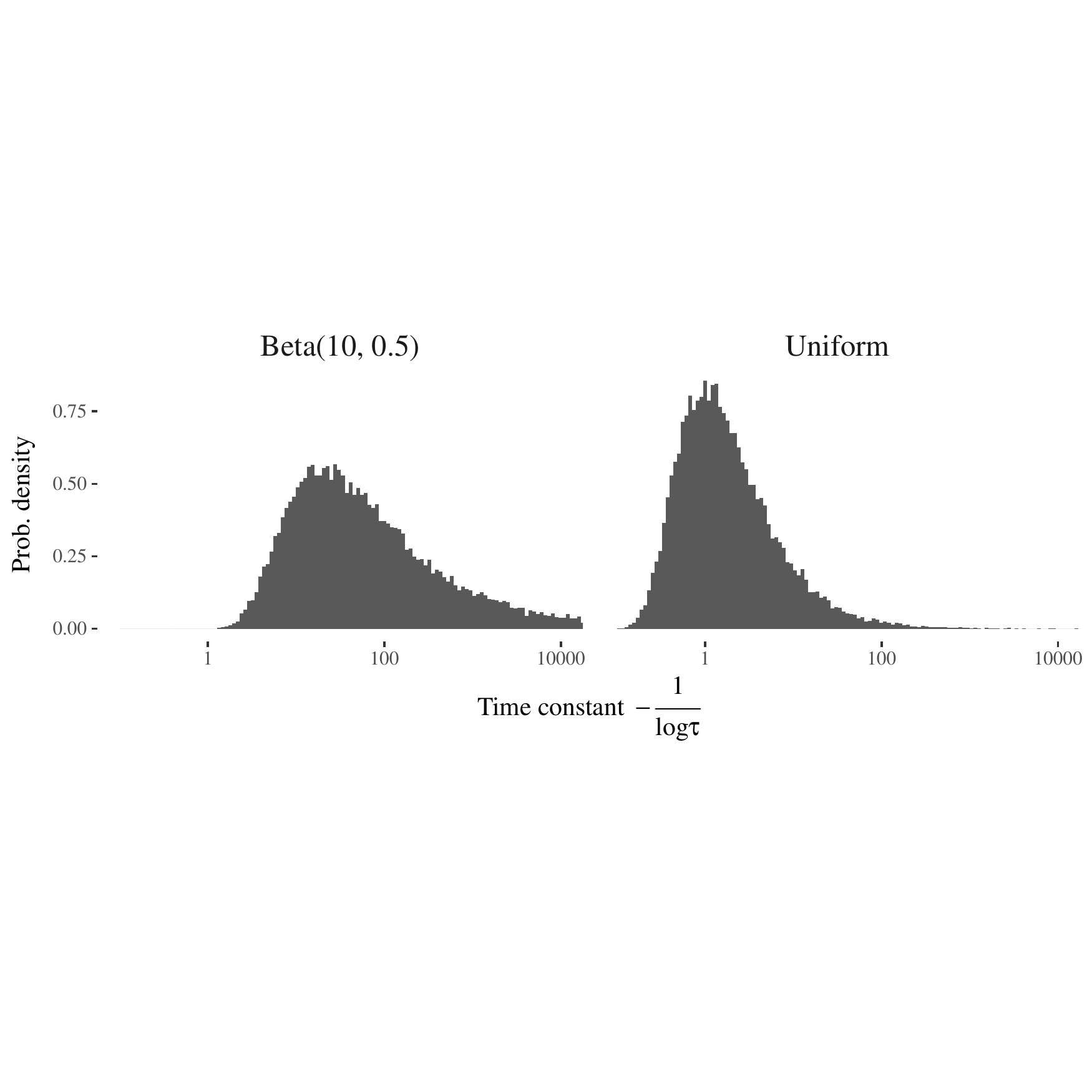}
  \caption{Comparison of $Uniform(0, 1)$ and $Beta(10, 0.5)$ prior for the
    parameter $\tau$. Each histogram consists of 25000 prior draws
    and shows the induced distribution on the corresponding time
    constant $- \frac{1}{\log \tau}$. Note that the uniform prior puts
    a considerable probability mass on very short time constants of
    below one time step. }
  \label{fig:tauPrior}
\end{figure}

Here, we propose the use of (weakly) informative priors which take
into account our knowledge about the role played by the parameters
when generating data from the likelihood model. As an
example, consider the parameter
$\tau \in (0, 1)$ of \eq{VS}. While it might be natural to simply
assign a uniform prior\footnote{Note that uniform priors, especially on unbounded spaces,
should not be considered as uninformative. On the one hand, an improper uniform prior, i.e.~when it cannot be normalized,
expresses a strong believe about extreme parameter values
by assigning infinite probability mass to values above any finite
threshold. On the other hand, they are not invariant under model
reparameterization as shown in the above example.}, $\tau$ controls the time constant of the
running price average, i.e.
\begin{eqnarray*}
  \langle p_{t} \rangle_{\tau} 
  & = & (1 - \tau) p_{t} + \tau \langle p_{t-1} \rangle_{\tau} \\
  & = & (1 - \tau) p_{t} + \tau \left( (1 - \tau) p_{t-1} + \tau \langle p_{t-2} \rangle_{\tau} \right) \\
  & = & (1 - \tau) \sum_{k = 0}^{\infty} \tau^k p_{t-k}
\end{eqnarray*}
as $\tau^k \to 0$ for $k \to \infty$. Comparing this to an exponentially weighted average in continuous
time, i.e.
\[ \langle p_{t} \rangle_{\rho} = \int_0^\infty \rho e^{- \rho s} p_{t
    - s} ds \] with time constant $\rho^{-1}$ and exponential
weighting kernel $k(s) = \rho e^{- \rho s}$ of unit weight, i.e.
$\int_0^\infty k(s) ds = 1$, we match $\tau^k$ with $e^{- \rho k}$.
Thus, interpreting $- \frac{1}{\log \tau}$ as the
time constant of the running average. \fig{tauPrior} compares an
$Uniform(0, 1)$ and a $Beta(10, 0.5)$ prior on $\tau$ in terms of the induced
prior on the time scale. Similarly, $\mu$ has been given a
$Gamma(3, 0.03)$ which assign more than $95\%$ of its probability mass
to the interval $[20, 250]$. Together these priors inform the VS model
to stay away from the boundary $\mu \to 0$ or $\tau \to 0$ where it becomes
trivial, i.e.~$\P(|S_t| = 1) \equiv 1$ and thus
$\sigma_t \equiv \sigma_{max}$. Accordingly, it cannot be expected to
generate data exhibiting pronounced volatility clustering in this case
and indeed, in the original reference \cite{VS}, prices where averaged
over $10^4$ time steps which is well covered by the chosen prior.
Implementing both models is straight-forward in {\em Stan} and the full code
can be found in appendix \ref{app:VS} and \ref{app:GARCH}
respectively.

\subsection{Model by Franke \& Westerhoff (FW)}
Franke \& Westerhoff have developed a series of models and estimated
using moment matching methods \cite{FWcontest,FWfit}. Here, we follow
their presentation in \cite{FWcontest} and introduce the DCA-HPM model
in their terminology.

In the FW model the market is populated with two types of agents namely fundamental
and chartist traders. The fraction of fundamental traders at time step
$t$ is denoted by $n_t^f \in [0, 1]$. The corresponding fraction of
chartist traders is then given by $n_t^c = 1 - n_t^f$. The log price,
denoted by $p_t$ adjusts to the average demand from fundamental $d^f$
and chartist $d^c$ traders as
\begin{equation}
  \label{eq:FWprice}
  p_t = p_{t-1} + \mu ( n_{t-1}^f d_{t-1}^f + n_{t-1}^c d_{t-1}^c )
\end{equation}
The demand is composed of a deterministic and stochastic component. It
is assumed that fundamental traders react to mispricing, i.e.~the
difference between $p_t$ and the (known) fundamental price $p^*$,
whereas chartist traders react to past price movement, i.e.
$p_t - p_{t-1}$. According to \cite{FWcontest} the demand dynamics is
modeled as
\begin{eqnarray*}
  d_t^f = \phi (p^* - p_t) + \epsilon^f_t && \epsilon^f_t \sim \mathcal{N}(0, \sigma_f^2) \\
  d_t^c = \xi (p_t - p_{t-1}) + \epsilon_t^c && \epsilon_t^c \sim \mathcal{N}(0, \sigma_c^2)
\end{eqnarray*}
Note that these demands are unobserved as only their weighted sum
effects the price. While such a dynamics could be modeled by means of
a stochastic latent state, in the present case it is possible to
marginalize out the demand. As the sum of two normally distributed
random variables is again normal, the combined demand gives rise to a
stochastic model for the log return $r_t = p_t - p_{t-1}$
\begin{equation}
  \label{eq:FWstoch_price}
  r_t \sim \mathcal{N}\left( \mu ( n_{t-1}^f \phi (p^* - p_{t-1}) + n_{t-1}^c \xi (p_{t-1} - p_{t-2}) ),
  \mu^2 ((n_{t-1}^f)^2 \sigma_f^2 + (n_{t-1}^c)^2 \sigma_c^2) \right)
\end{equation}
The volatility
$\sigma_t = \mu \sqrt{(n_{t-1}^f)^2 \sigma_f^2 + (n_{t-1}^c)^2
  \sigma_c^2}$
now depends on the fraction of chartist vs fundamental traders and
changes over time. \cite{FWcontest} calls this structured stochastic
volatility, in analogy to structural models in economics, as the
parameters of the agent-based model are grounded in behavioral terms
and therefore economically meaningful.

The model is then completed by an update equation for the fraction of
traders in each group. Here, we consider the DCA-HPM specification of
\cite{FWcontest} which is given by
\begin{eqnarray}
  \label{eq:FW_DCA}
  n_t^f & = & \frac{1}{1 + e^{- \beta a_{t-1}}} \\
  n_t^c & = & 1 - n_t^f \nonumber \\
  \label{eq:FW_HPM}
  a_t & = & \alpha_0 + \alpha_n (n_t^f - n_t^c) + \alpha_p (p^* - p_t)^2
\end{eqnarray}
The parameter $a_t$ denotes the relative attractiveness of the
fundamental over the chartist strategy. It includes a general
predisposition $\alpha_0$ and herding $\alpha_n > 0$ as well as
mispricing $\alpha_p > 0$ effects. We chose this specification for
two reasons:
\begin{enumerate}
\item The discrete choice approach (DCA) of \eq{FW_DCA} leads to a
  smoothly differentiable model density. This eases the exploration
  of the posterior when sampling with the HMC algorithm.
\item The herding $+$ predisposition $+$ misalignment (HPM)
  specification for the attractiveness \eq{FW_HPM} can be computed
  without access to the actual demands $d_t^f$ and $d_t^c$. This is
  not true for the other specifications of \cite{FWcontest} where the agents
  wealth depends on previous demands which, in turn, leads to a stochastic
  volatility model where (one of) the demands has to be modeled as a
  stochastic latent variable. For simplicity we have not considered
  this complication in the present paper.
\end{enumerate}
Overall, the model dynamics is fully specified by \eq{FWstoch_price},
\eq{FW_DCA} and \eq{FW_HPM}. The parameters of the model are given by
$\vt^{FW} = (\mu, \phi, \sigma_f, \xi, \sigma_c, \beta, \alpha_0,
\alpha_n, \alpha_p, p^*)$.
Note that $\beta$ and $\mu$ are redundant as they simply control the
scale of $\alpha_0, \alpha_n, \alpha_p$ and
$\xi, \phi, \sigma_f, \sigma_c$ respectively. Thus, throughout we fix
them at $\beta = 1$ and $\mu = 0.01$ as in the simulation exercise of
\cite{FWcontest}.

When simulating data we further assume that the fundamental log price
is known and fixed at $p^* = 0$. When estimating the model on real
stock returns below, we do not know the fundamental price. In this
case, following \cite{LuxSMC}, we assume that the log fundamental
price is time varying as a Brownian motion
\[ p^*_t \sim \mathcal{N}(p^*_{t-1}, \sigma_*^2) \]
This not only introduces another parameter $\sigma_*$ but also renders
the model a stochastic volatility model, i.e.~the volatility
$\sigma_t$ now includes a stochastic component. To see this note that
$\sigma_t$ depends on $a_{t-2}$ via $n_{t-1}^f$ and the attractiveness
in turn includes the stochastic fundamental log price $p^*_{t-2}$.

Nevertheless, implementing the model in {\em Stan} is readily possible. As
before, the full code of the FW model is given in appendix
\ref{app:FW}. Note that the time varying noise of the Brownian motion
$p^*_t$ appears as an $N$-dimensional vector (where $N$ denotes the
number of observed time steps) in the parameter block. In addition, we
have used a non-centered parameterization,
i.e.~\lstinline[style=customInline]{p_star} is computed from
\lstinline[style=customInline]{epsilon_star} as a transformed
parameter. Formally, we can express this as follows:
\[ p^*_t = p^*_{t-1} + \sigma_* \epsilon^*_t \quad \mbox{where} \quad
  \epsilon^*_t \sim \mathcal{N}(0, 1) \]
instead of
\[ p^*_t \sim \mathcal{N}(p^*_{t-1}, \sigma_*^2) \]
This is a standard example of a reparameterization which does not
change the model, but helps when HMC sampling as the raw parameters
$\epsilon_t^*$ all have unit scale, no matter which variance
$\sigma_*^2$ is currently sampled.

Here, we complete the model with weakly informative priors for
all parameters. As few insights
are available about the proper choice of the attractiveness parameters $\alpha_0, \alpha_n$ and $\alpha_p$ we assign weakly informative priors, e.g.
\lstinline[style=customInline]{alpha_0 ~ student_t(5, 0, 1)}, which restrict
the scale of the parameter yet, being heavy tailed, allow substantially larger values\footnote{In contrast, a normal prior distribution
  would impose much more information as values larger than several
  tens of standard deviations are essentially ruled out.}. In case of the standard deviation parameters
$\sigma_f, \sigma_c$ and $\sigma_*$ we impose stronger priors and resort to the observed
data to set the proper scale. While not being purely Bayesian, this choice restricts the model to
reasonable scales accounting for the fact that volatility could be measured in arbitrary units, e.g.~percent per
year. Overall, we found these priors effective in simulation studies
as well as when estimating the model on stock data.

\section{Results}
Here, we present estimation results of the above models, both
on simulated and on real price data.

\subsection{Simulation studies}
\begin{figure}[ht]
  \centering
  \includegraphics[width=1\textwidth, trim=0 80 0 80]{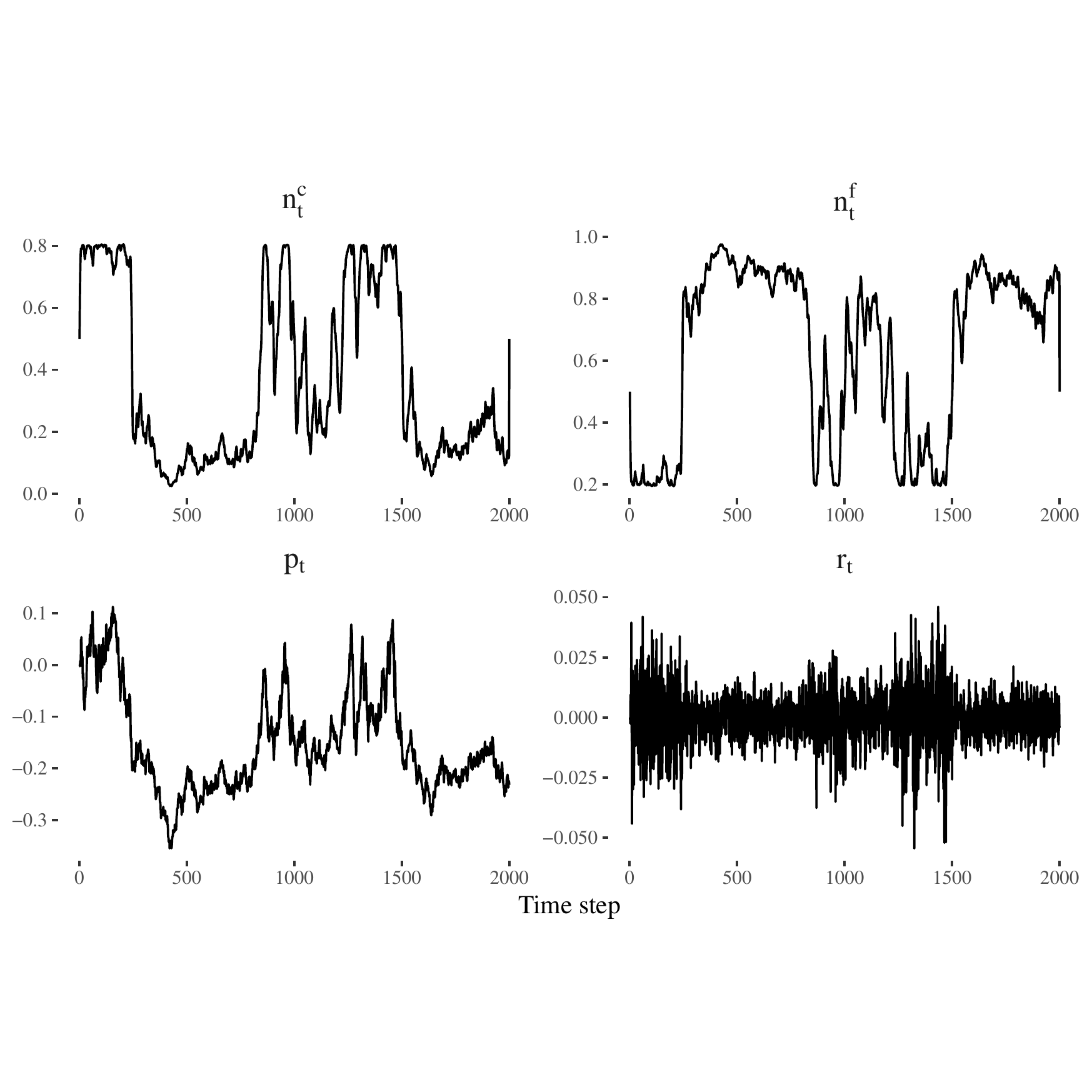}
  \caption{Simulated price and return series of FW model. Note that a
    low fraction of fundamental traders $n_f$ coincides with volatile
    market phases.}
  \label{fig:FWsim}
\end{figure}

In order to check our model implementation, we simulated the FW model
with parameters as given in table 1 of \cite{FWcontest}, i.e.
$\mu = 0.01, \beta = 1, \phi = 0.12, \xi = 1.50, \alpha_0 = -0.327,
\alpha_n = 1.79, \alpha_p = 18.43, \sigma_f = 0.758, \sigma_c = 2.087$
and $p^* = 0$. Then, we re-estimated the model parameters\footnote{For
  this exercise, we have adapted the model code of app.~\ref{app:FW}
  such that the fundamental price is fixed at $p^* \equiv 0$. This
  matches the data generating process as well as the calibration
  exercise in \cite{FWcontest}.} on the simulated price series of
$T = 2000$ time steps shown in \fig{FWsim}.

\begin{figure}[ht]
  \centering
  \includegraphics[width=1\textwidth, trim=0 80 0 80]{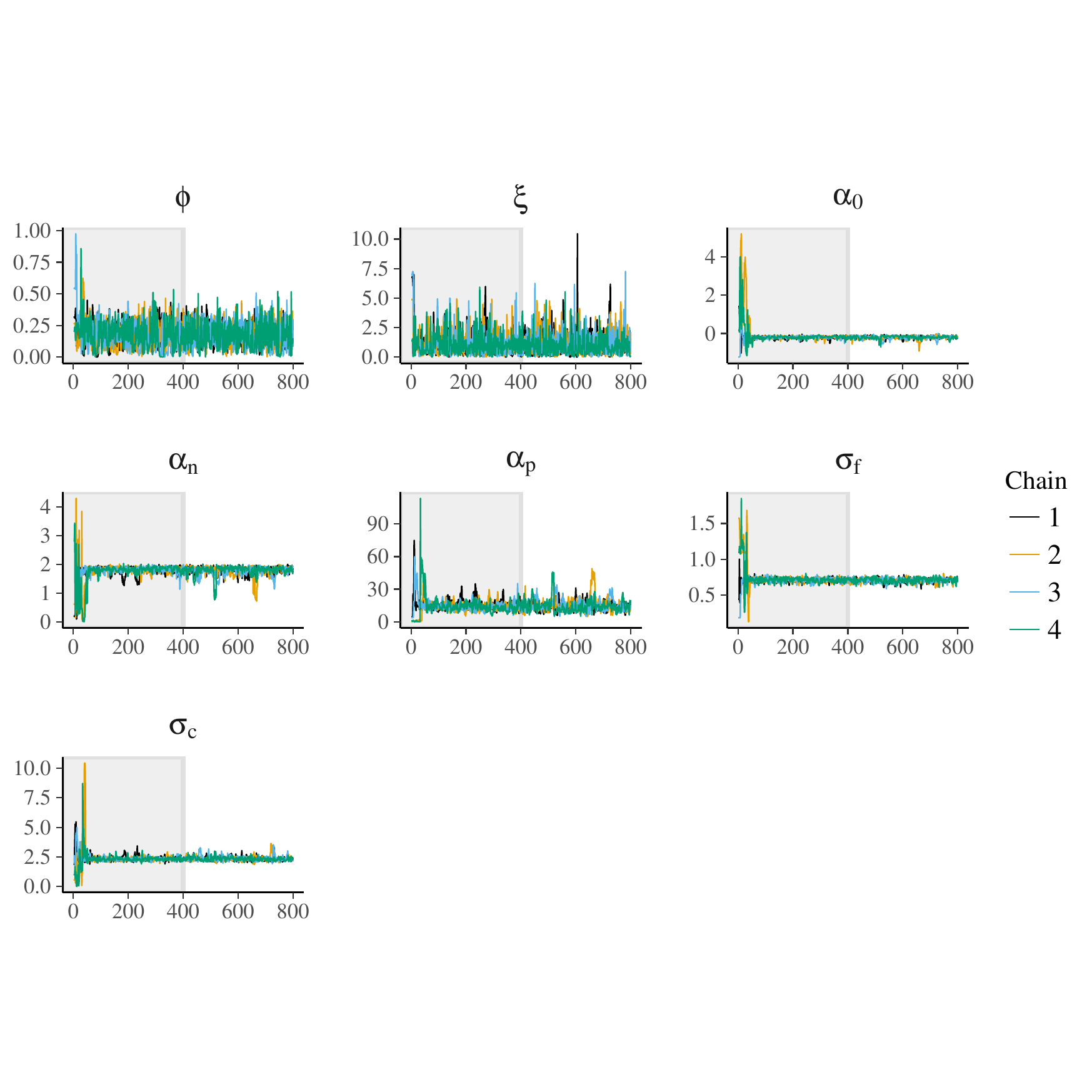}
  \caption{Trace plot for model parameters
    $\phi, \xi, \alpha_0, \alpha_n, \alpha_p, \sigma_f$ and
    $\sigma_c$. Note that all chains appear to have converged to the
    same posterior distribution after just about 50 samples.}
  \label{fig:FWsimfit_trace}
\end{figure}

\begin{figure}[ht]
  \centering
  \includegraphics[width=1\textwidth, trim=0 80 0 80]{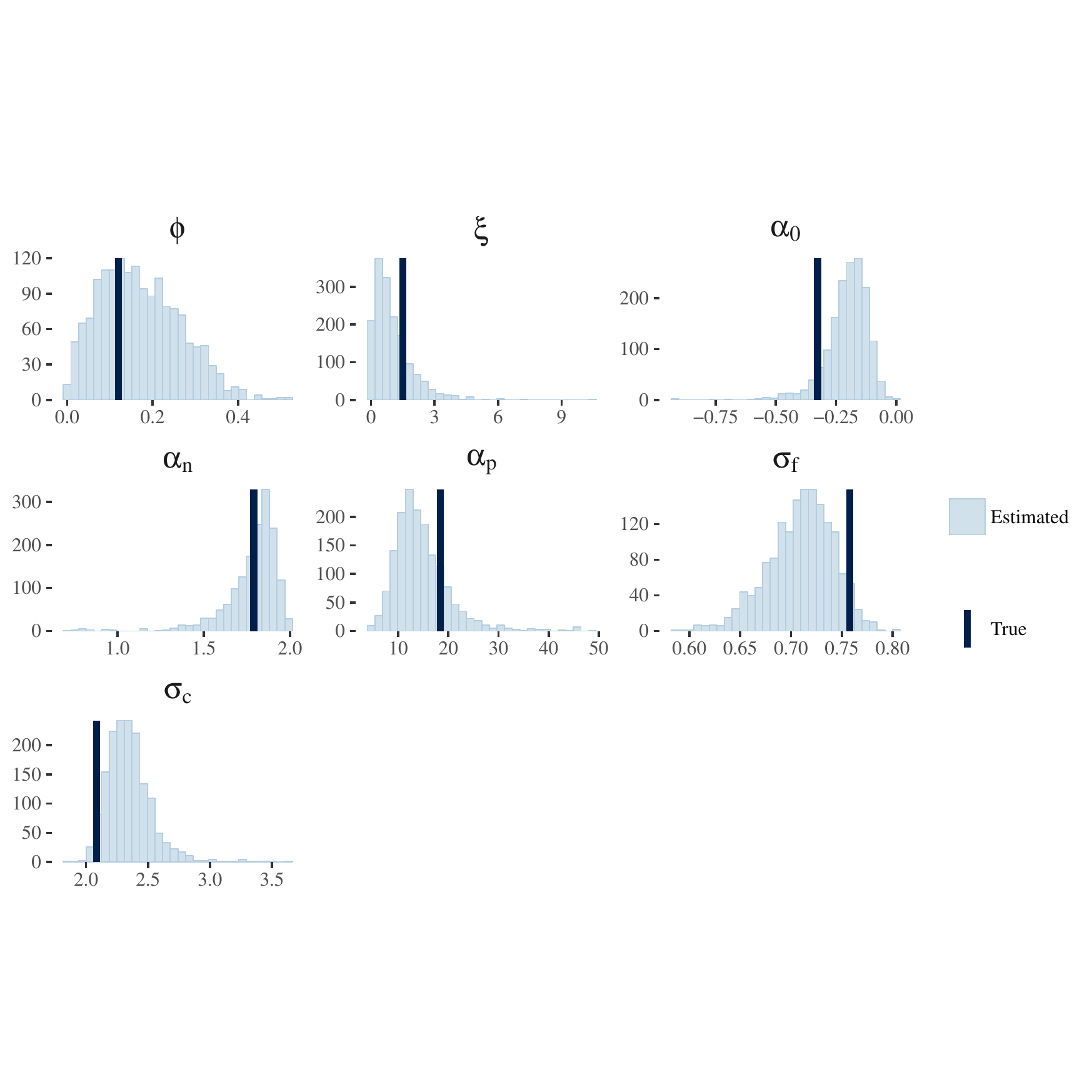}
  \caption{Plot of posterior densities for parameters
    $\phi, \xi, \alpha_0, \alpha_n, \alpha_p, \sigma_f$ and
    $\sigma_c$. The true values are well covered by the posterior
    distributions.}
  \label{fig:FWsimfit_recover}
\end{figure}

The resulting samples from the posterior distribution are shown in
\fig{FWsimfit_trace}. Overall, we have run four chains
starting from independent random initial conditions and drawn 400
samples from each after discarding an initial transient of another 400
samples as warmup. Compared to other studies the number of samples
appears very low, but the high quality of the samples is clearly
visible in the trace plots. The model appears to have converged after
just about 50 samples and all chains produce almost uncorrelated
samples from the same distribution. This is also confirmed by standard
convergence diagnostics such as Gelman \& Rubin's $\hat{R}$, which compares the
variance between and within chains, or the number of effective
samples, which is based on the sample auto-correlation (not shown). If desired,
more samples can easily be drawn as the shown estimation runs in a few
minutes on a standard laptop.
\fig{FWsimfit_recover} shows the resulting posterior
distributions together with the true parameters that generated the
data. We found that about at least 1000 observed prices are
necessary to reliably recover the true parameters. Interestingly,
preliminary runs on considerably longer time series of 5000 observations
suggest that posterior uncertainty reduces only slightly, especially of the
chartist parameters which appear harder to estimate, presumably because they become
effective during episodes of high volatility only.

Similar experiments have been carried out with the VS model. Here, all
parameters are precisely estimated after a few thousand
observations. Especially $\tau$ has very small uncertainty while $\mu$
appears to be the least well identified parameter. A summary of our
experiments for the FW and VS models can be found in table
\ref{tab:recover}. Overall, we simulated 100 time series of 2000 time
steps each and took the posterior means as point estimates for
parameters.
\begin{table}[ht]
  \begin{subtable}{1\textwidth}
    \caption{FW model estimation results}
    \begin{tabular}{llllllll}
      Parameter & $\phi$ & $\xi$ & $\sigma_f$ & $\sigma_c$ & $\alpha_0$ & $\alpha_n$ & $\alpha_p$ \\ \hline
      True & 0.12 & 1.5 & 0.758 & 2.087 & $-0.327$ & 1.79 & 18.43 \\ \hline
      Estimates & & & & & & & \\ \hline
      Mean & 0.23 & 0.97 & 0.75 & 2.14 & $-0.28$ & 1.83 & 16.9 \\
      SD   & 0.23 & 0.18 & 0.056 & 0.19 & 0.14 & 0.22 & 6.0 \\
      RMSE & 0.26 & 0.55 & 0.056 & 0.20 & 0.14 & 0.22 & 6.2 \\ \hline
    \end{tabular}
  \end{subtable} \\[2ex]
  \begin{subtable}{1\textwidth}
    \caption{VS model estimation results}
    \begin{tabular}{llll}
      Parameter & $\mu$ & $\tau$ & $\sigma_{max}$ \\ \hline
      True & 100 & 0.999 & 0.01 \\ \hline
      Estimates & & & \\
      Mean & 95.9 & 0.997 & 0.011 \\
      SD & 11.9 & $7.8 \times 10^{-3}$ & $4.0 \times 10^{-3}$ \\
      RMSE & 12.6 & $8.0 \times 10^{-3}$ & $4.2 \times 10^{-3}$ \\ \hline
    \end{tabular}
  \end{subtable}%
  \caption{\label{tab:recover} Estimation results on simulated data for the FW (a) and VS (b) model. Shown are the mean, standard deviation (SD) and root mean squared errors (RMSE) of posterior mean point estimates over 100 simulated time series of 2000 time steps each.}
\end{table}

\begin{figure}[ht]
  \centering
  \includegraphics[width=1\textwidth, trim=0 80 0 80]{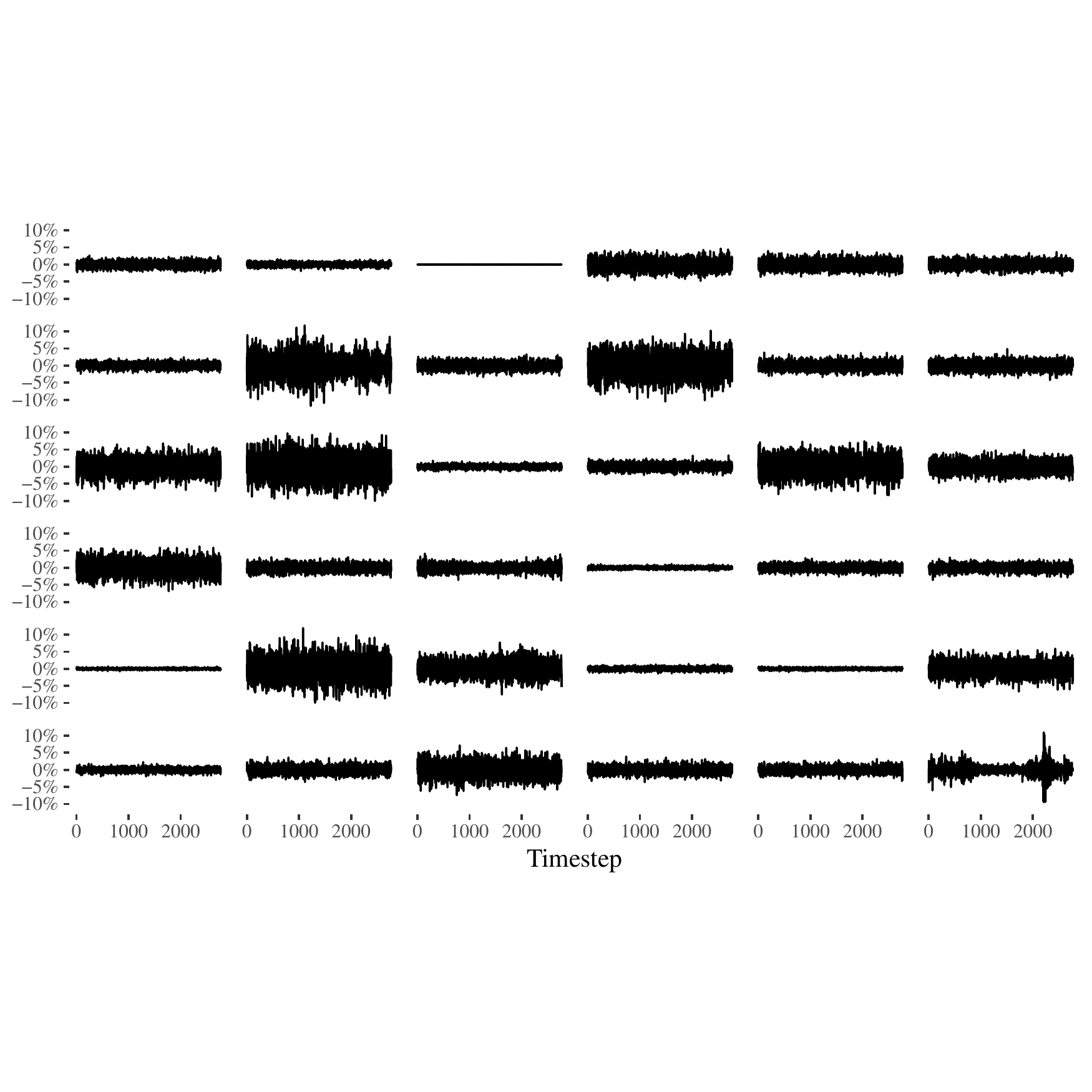}
  \caption{Return series generated from the FW model when parameters
    are drawn according to the prior. For comparison actual returns of
    the S\&P 500 are included in the lower right sub panel.}
  \label{fig:FWprior}
\end{figure}

The above exercises of parameter re-estimation are based on the idea
that actual data had been generated from a fixed, yet unknown, set of
parameters that is to be recovered. This is deeply routed in
frequentist statistics where the properties of an estimator are
compared by its repeated sampling properties, i.e.~investigating its
performance on several data sets generated from the same underlying
true data generating process such as a model with fixed parameters. From a Bayesian perspective, as well as from a data
modeling point of view, inference should instead be based on the
observed data set alone. As explained above, a Bayesian model consists
of a joint distribution on data and parameters with density
$p(\vx, \vt)$. The corresponding (prior)
distribution on data $p(\vx)$ then captures our modeling assumption about
which observations are considered plausible. Thus, from this perspective
it is natural to investigate properties of this distribution and
\fig{FWprior} shows randomly generated return series 
from the FW model with our chosen priors.
For comparison actual S\&P 500 returns are included as the lower right
sub panel. It is clearly visible that volatility clustering under the
model is much less pronounced. Investigating different prior
specifications, in particular trying to obtain stronger volatility
clustering, reveals an intricate relationship between the scale of the
volatility parameters $\sigma_f, \sigma_c$ and the typical size of the
choice parameters $\alpha_n, \alpha_p$. Indeed, looking at \eq{FW_HPM}
this should not come as a surprise, as $\sigma_c$ and $\sigma_f$
control the expected price fluctuations which in turn set a scale for
$\alpha_n$ and (less directly) for $\alpha_p$. Overall, this data
generating exercise not just aids in understanding the FW model and
its assumption, but also suggests that $\alpha_n, \alpha_p$ would be
easier interpretable if considered on a relative scale,
e.g.~reparameterized as
$\widetilde{\alpha}_n = \frac{\alpha_n}{\langle \sigma_t \rangle}$
where $\langle \sigma_t \rangle$ denotes the expected
volatility. Leaving such explorations for future work, we now turn to
fitting the models on stock returns of the S\&P 500 index.

\subsection{Fitting the S\&P 500}

\begin{figure}[ht]
  \centering
  \includegraphics[width=0.9\textwidth, trim=0 80 0 80]{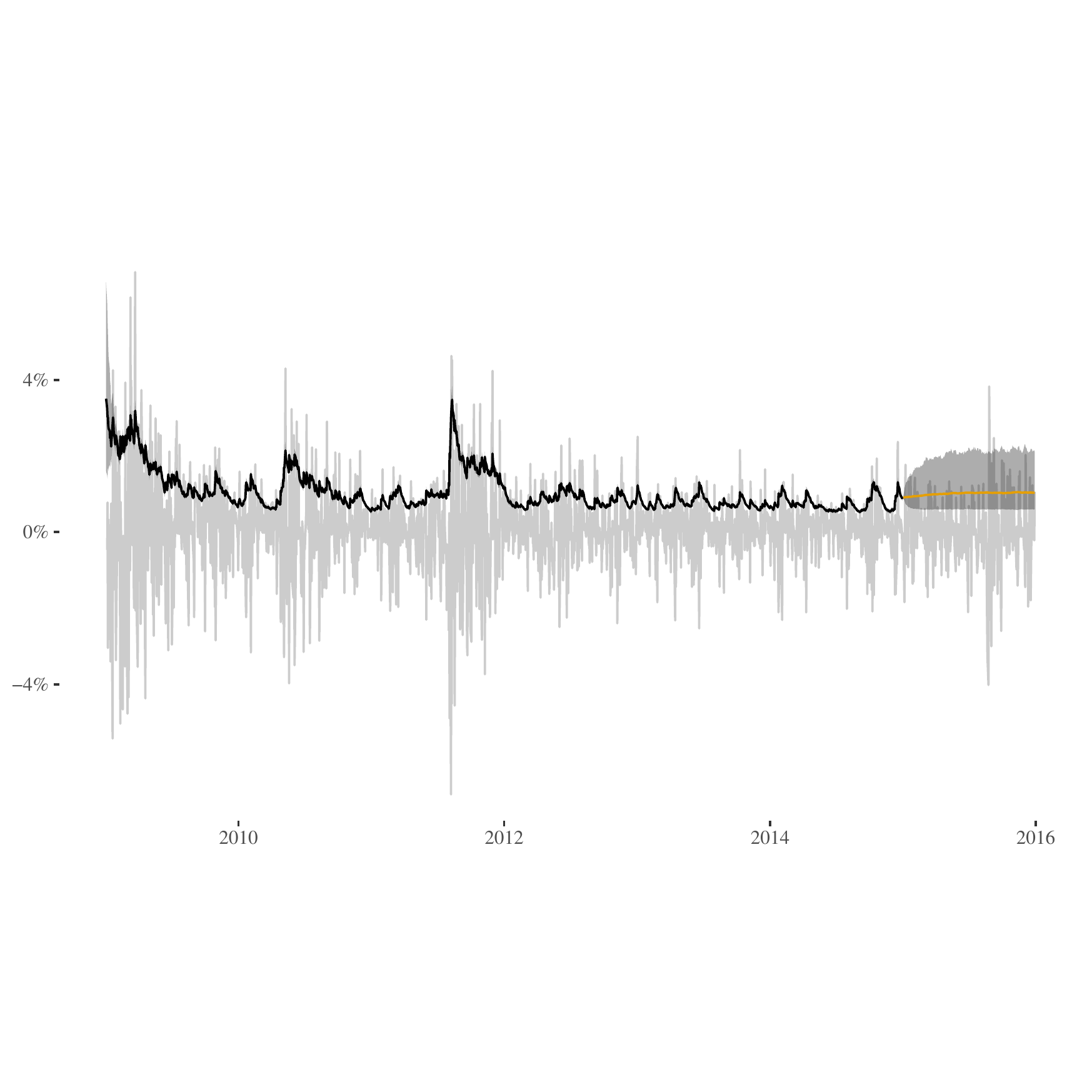}
  \includegraphics[width=0.9\textwidth, trim=0 80 0 80]{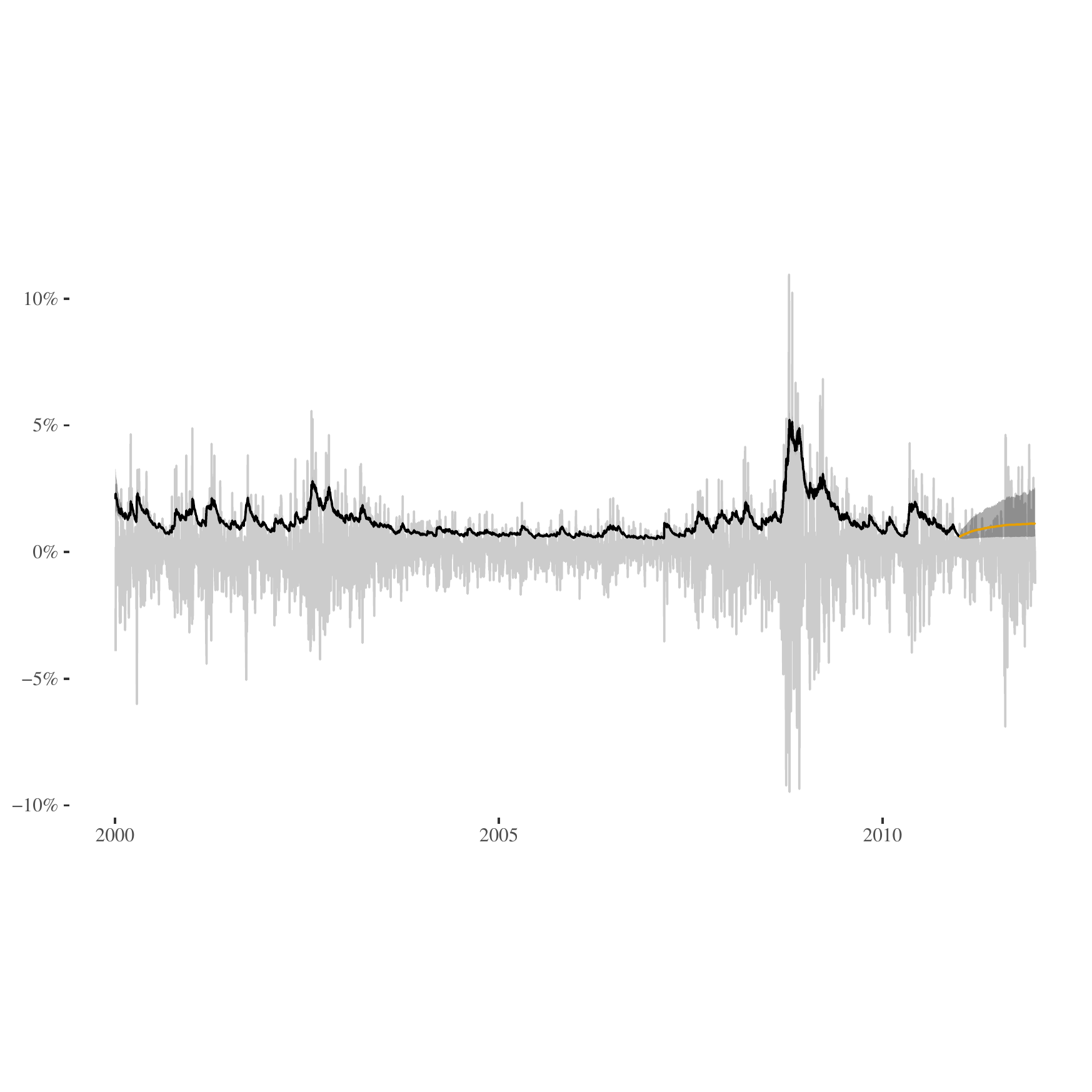}
  \caption{GARCH model fit and predictions on the S\&P 500.}
  \label{fig:GARCHfitSP500}
\end{figure}

\begin{figure}[ht]
  \centering
  \includegraphics[width=0.9\textwidth, trim=0 80 0 80]{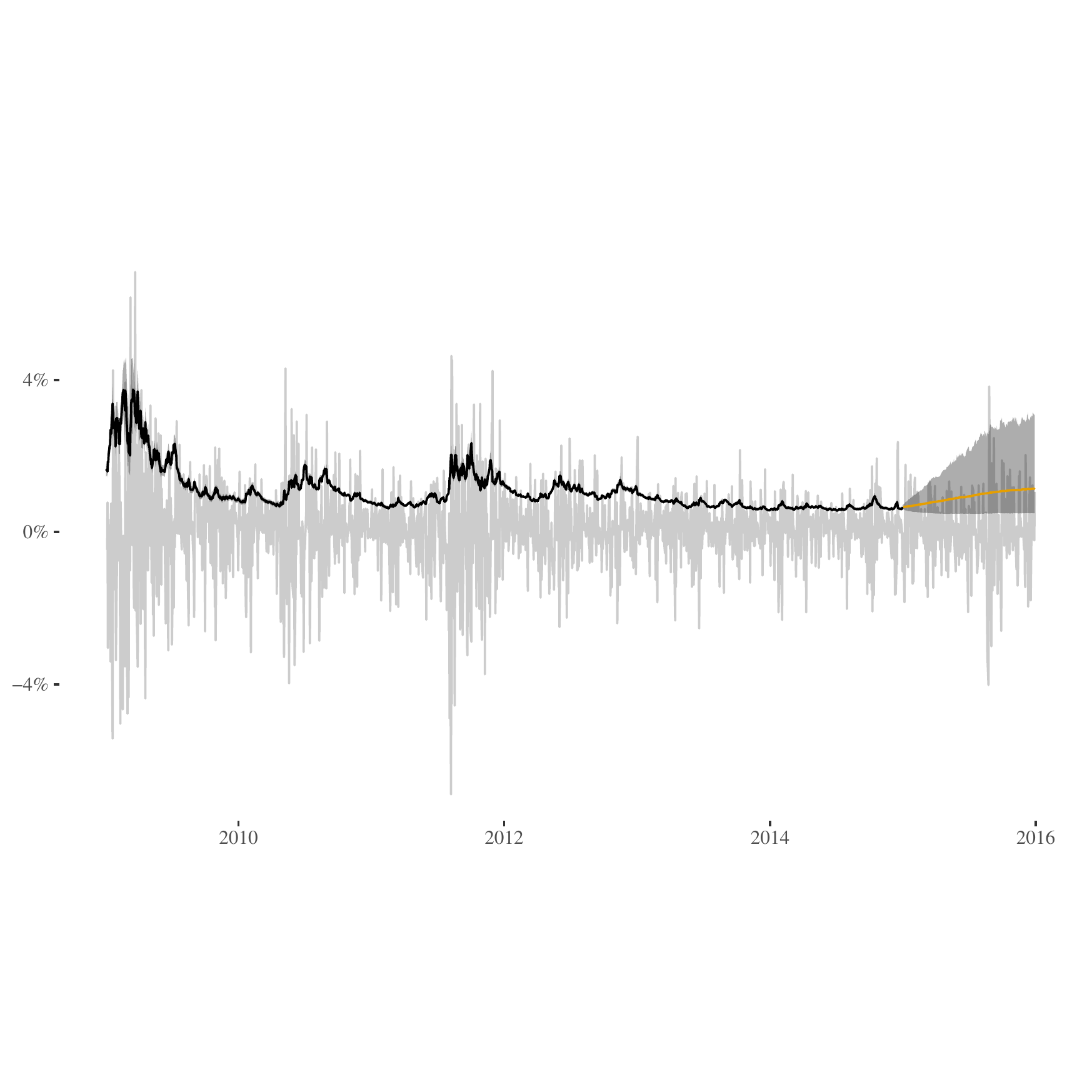}
  \includegraphics[width=0.9\textwidth, trim=0 80 0 80]{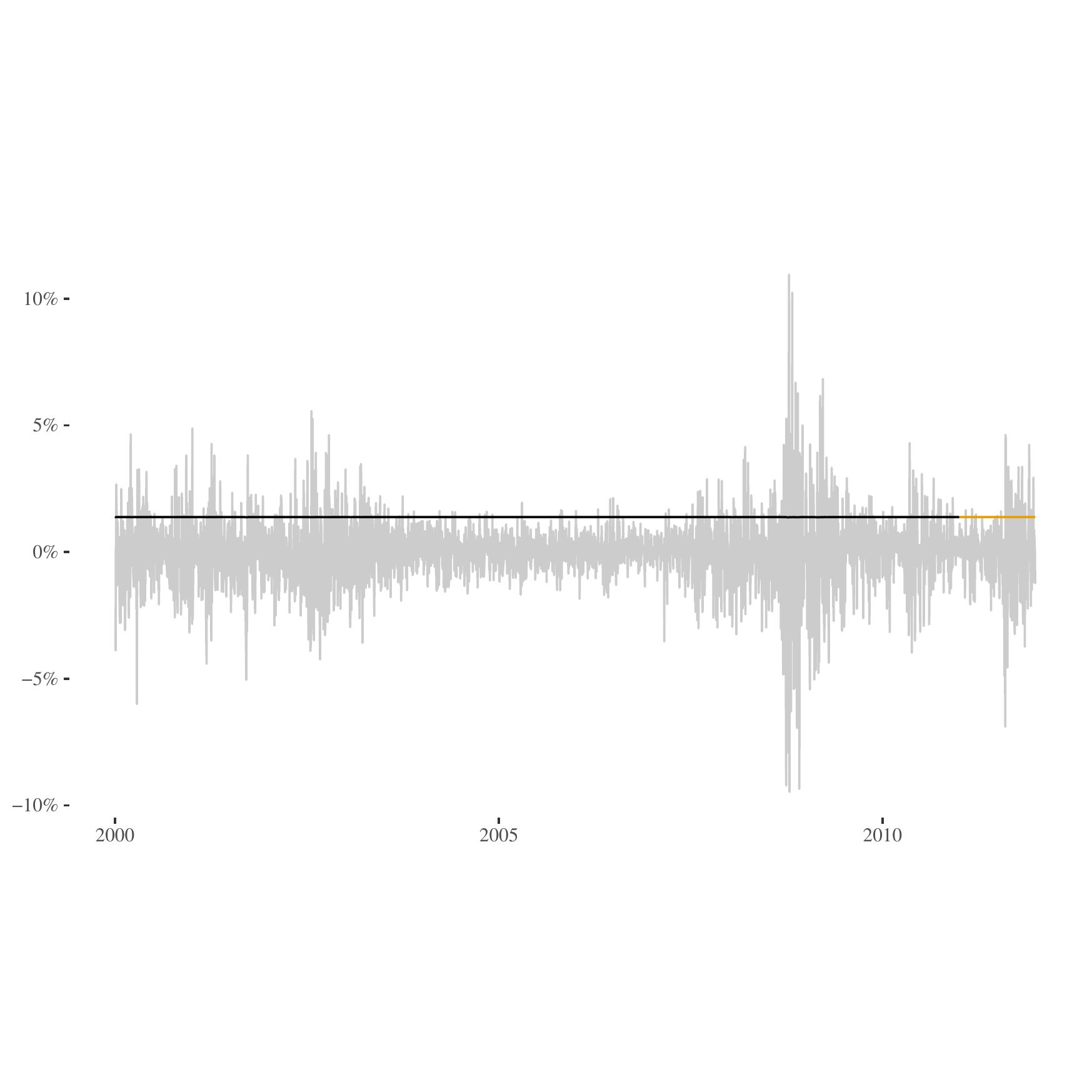}
  \caption{VS model fit and predictions on the S\&P 500.}
  \label{fig:VSfitSP500}
\end{figure}

\begin{figure}[ht]
  \centering
  \includegraphics[width=0.9\textwidth, trim=0 80 0 80]{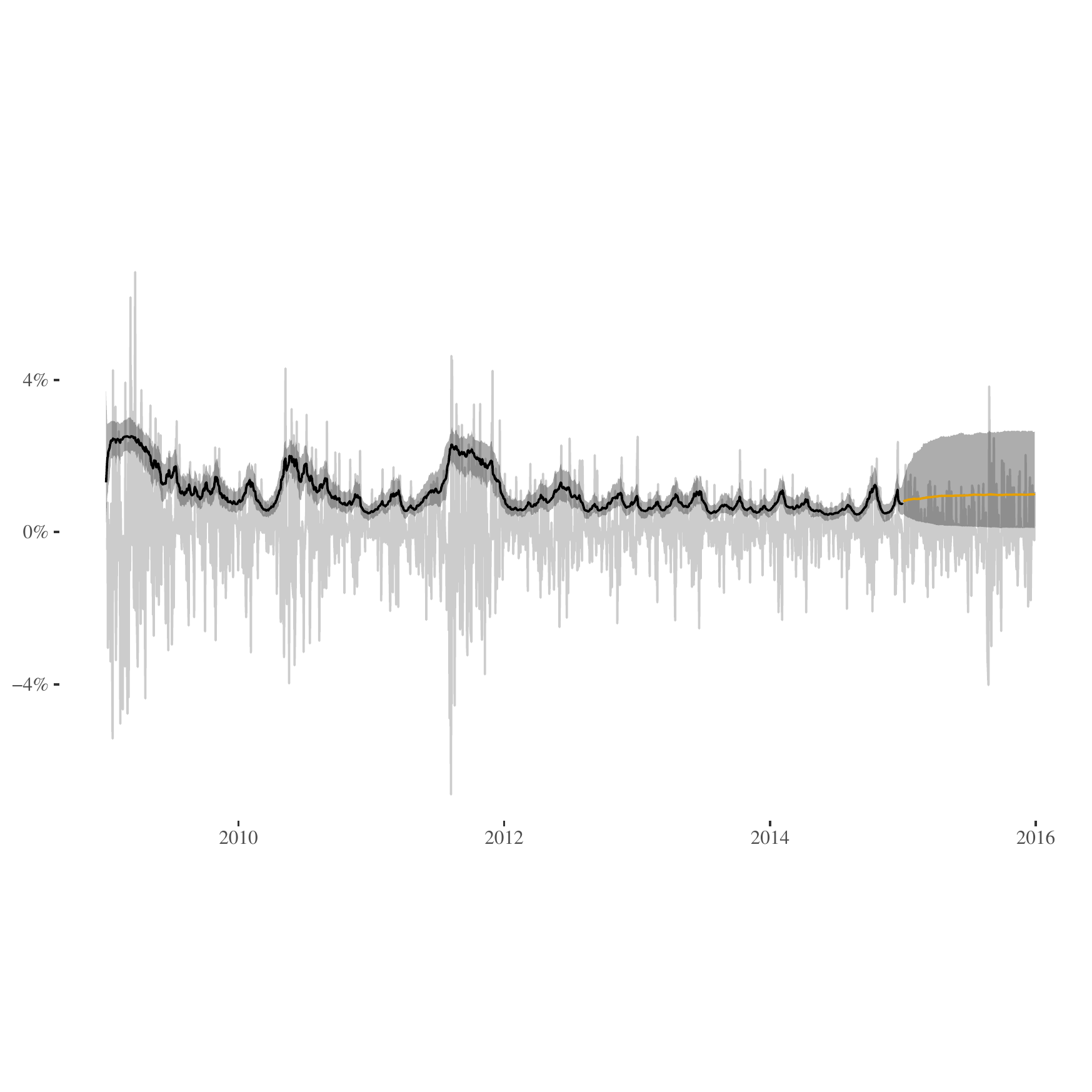}
  \includegraphics[width=0.9\textwidth, trim=0 80 0 80]{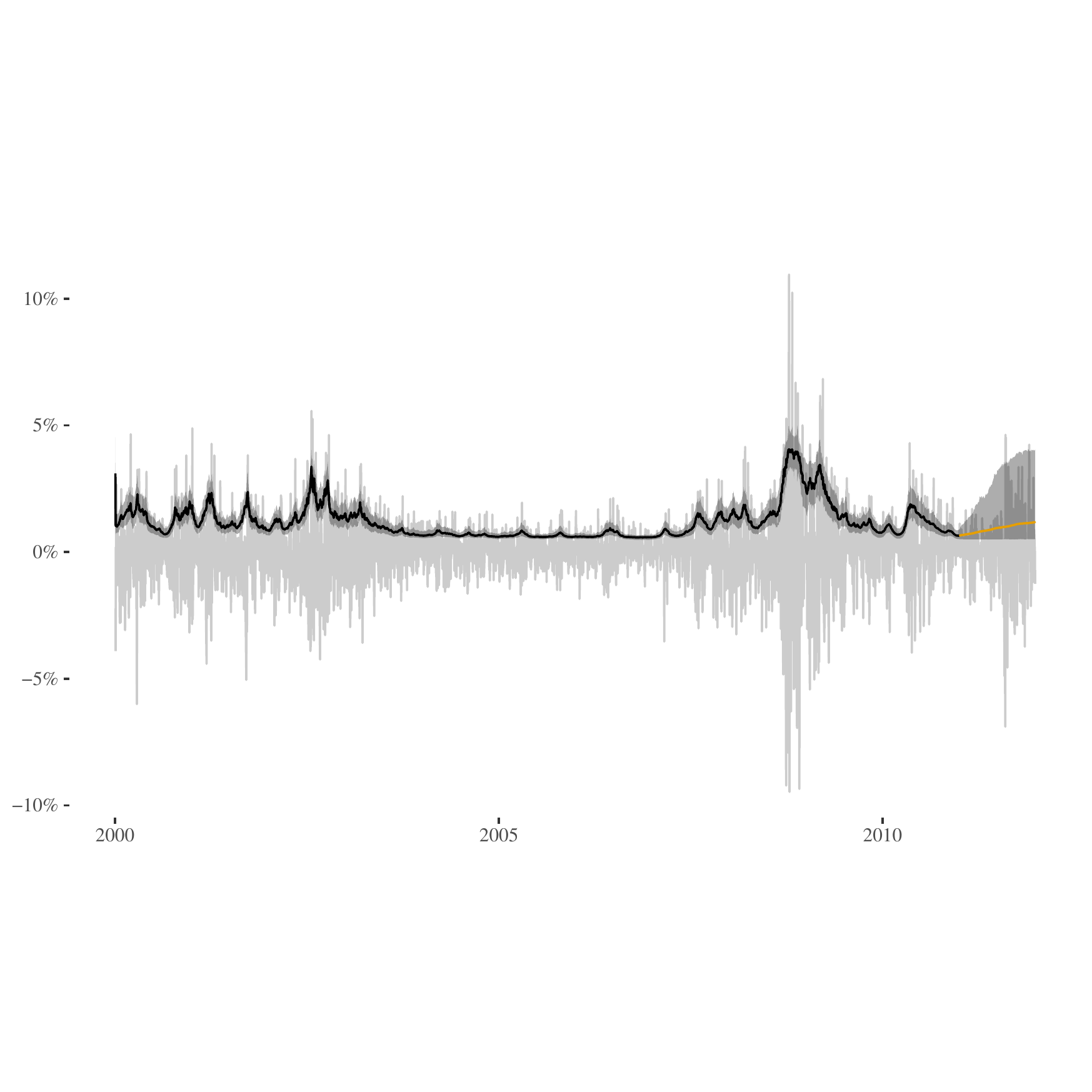}
  \caption{FW model fit and predictions on the S\&P 500.}
  \label{fig:FWfitSP500}
\end{figure}

Finally, we have fitted all models on price data from the S\&P 500
stock market index. As a benchmark, a standard GARCH(1, 1) model has
been included for comparison. The corresponding fits from Jan 2009 to
Dec 2014 and Jan 2000 to Dec 2010 are shown in \fig{GARCHfitSP500},
\fig{VSfitSP500} and \fig{FWfitSP500}. The estimated model volatility
as well as prediction 250 days ahead is overlaid on the actual market
returns.

Volatility estimates are shown as the posterior mean together with the
$95\%$ credibility bands around it. The posterior of the volatility
$\sigma_i$ at time step $i$ is based on data points from returns $r_i$
observed over all $N$ data points, i.e.
$p(\sigma_i | r_1, \ldots, r_N)$ for $i = 1, \ldots, N$. In the
terminology of time series models this is known as the smoothing
distribution, in contrast to the filtering distribution
$p(\sigma_i | r_1, \ldots, r_{i-1})$ which is conditioned on previous
observations only. In this respect, our results complement the work of
Lux \cite{LuxSMC} who has used sequential Monte-Carlo to
approximate the filtering distribution. In the context of time series,
it is often convenient to evaluate models based on rolling look-ahead
predictions. While these are readily available from the filtering
distribution more work and successive re-fitting of the model would be
required in our setup. In any case, predictions beyond the last
observed data point are derived in the same fashion from the posterior
predictive distribution with density
\[ p(r_{N+1}, \ldots | r_1, \ldots, r_N) = \int p(r_{N+1}, \ldots |
\vt) p(\vt | r_1, \ldots, r_N) d\vt, \]
i.e.~by running the model forward based on the posterior
distribution. Note that the unobserved state $\vt$ includes parameters
as well as time varying states of the model, e.g.~the instantaneous
volatility $\sigma_1, \ldots, \sigma_N$ or the fraction of chartist
traders $n^c_1, \ldots, n^c_N$, rendering the predictions independent
from past observations.

Comparing the volatility estimates and predictions of the different
models, a few remarks are in order:
\begin{itemize}
\item The VS model appears to be severely misspecified. Quite often,
  especially if the chosen time window did not start out at high
  volatility, the best fit consists of constant
  volatility\footnote{Despite the chosen informed priors striving to
    avoid this trivial solution.}. Overall, the model appears unable
  to match the shape of empirical volatility clusters.
\item The vanilla GARCH(1,1) model provides reasonable volatility
  estimates. Albeit the small uncertainty when predicting
  ahead suggests that it cannot generate the heavy tails of empirical
  returns\footnote{Of course, this is well known and addressed by many
  extensions of the model.}. It still provides a reasonable benchmark to
  compare the presented agent-based models against.
\item Overall, the FW seems to fit the data best. In particular, the
  wider prediction intervals indicate its ability to generate heavy
  tailed return series. Due to the stochastic volatility process arising from
  the Brownian motion modeling the unobserved fundamental price,
  uncertainty of its volatility estimates is higher than for the other
  models. This feature is shared with other, more standard stochastic
  volatility models \cite{StochVol}.
\end{itemize}

Having compared the models graphically, we proceed with a more
principled assessment of model fit. In general, the predictive
likelihood will be biased upwards when evaluated on the data used to
estimate parameters and, in the case of nested models, prefers the
more complex variants as these cannot fit the data worse than
restricted variants. Thus, a fair assessment should address the models
ability to predict new data. In a time series context it is rather
natural to consider (running) look-ahead predictions. Here, for
computational reasons, we instead resort to leave-one-out (LOO)
predictions, i.e.~predicting current returns in the context of past
and future returns, excluding the current one. With the method of
Pareto smoothed importance sampling (PSIS) \cite{PSIS} the
corresponding predictive likelihoods
$p(r_i | r_1, \ldots, r_{i-1}, r_{i+1}, \ldots, r_N)$ can be estimated
from posterior samples. Note that in contrast to the full posterior
$p(\vt | \bm{r})$, the LOO likelihood is conditioned on all but the
$i$th data point. Thus, in order to estimate the LOO likelihood from
posterior samples the $i$th data points needs to be effectively
removed before evaluating the prediction. In general, this is far from
trivial and we refer the reader to \cite{PSIS} for details about how
PSIS estimates LOO likelihood. Here, we simply give the results of
model comparison in table \ref{tab:compare}. They confirm our previous
analysis, that the FW model fits the data best followed by the GARCH
model. The VS model seems worst even when estimating non-constant volatility.
\begin{table}[ht]
  \centering
  \begin{tabular}{lll}
    & \multicolumn{2}{c}{Time window} \\ \hline
    Model & Jan. 2009 -- Dec. 2014 & Jan. 2000 -- Dec. 2010 \\ \hline
    GARCH & $4868 \pm 37$ & $8574 \pm 53$ \\
    VS & $4845 \pm 40$ & $7921 \pm 82$ \\
    FW & $4916 \pm 37$ & $8641 \pm 49$
  \end{tabular}
  \caption{Model comparison based on leave-one-out predictive likelihoods.}
  \label{tab:compare}
\end{table}

How does the FW model achieve such a good fit of the time varying
volatility? This is particularly interesting, as its parameters are
interpretable in behavioral terms.

\begin{figure}[ht]
  \centering
  \includegraphics[width=1\textwidth, trim=0 80 0 80]{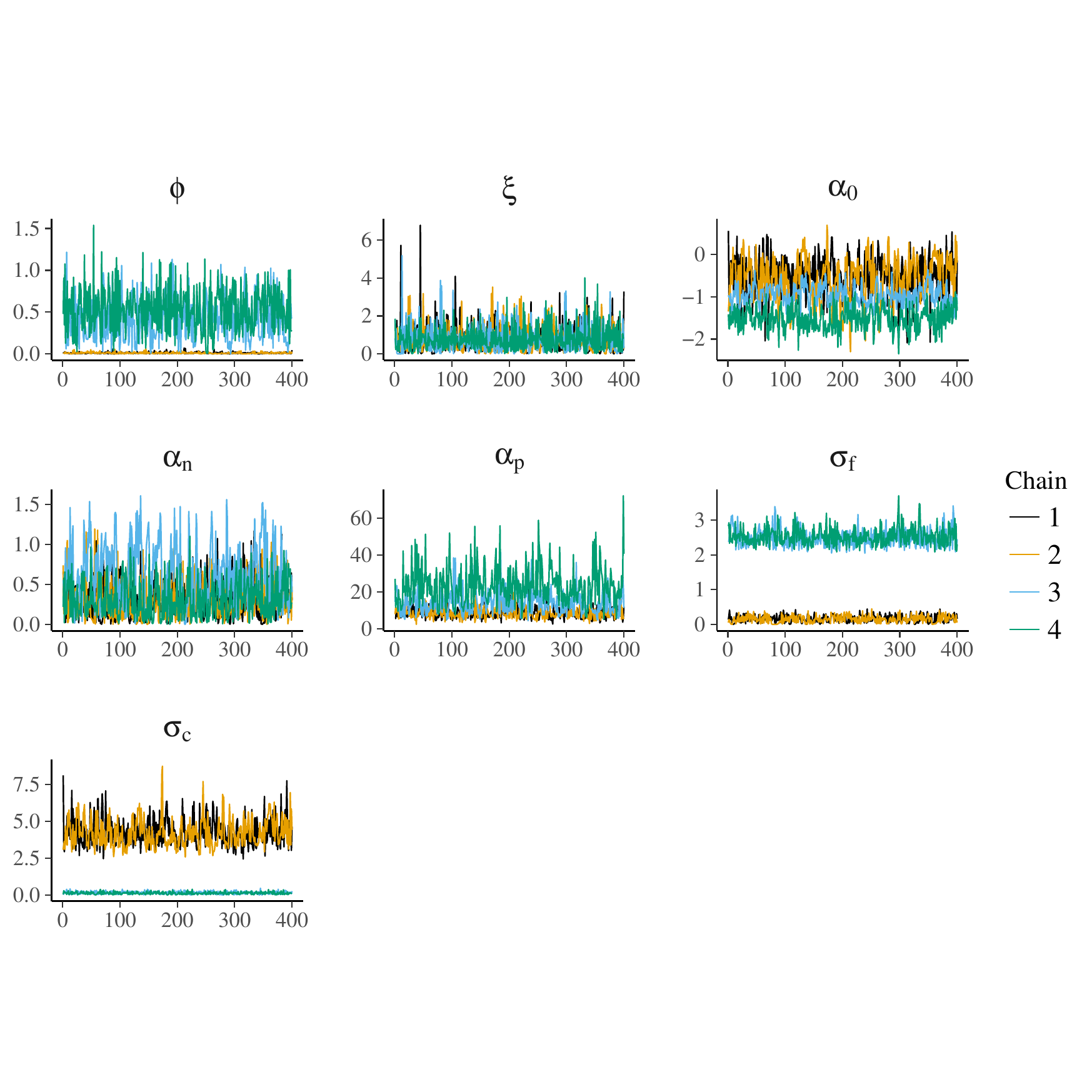} \\
  \includegraphics[width=1\textwidth, trim=0 80 0 80]{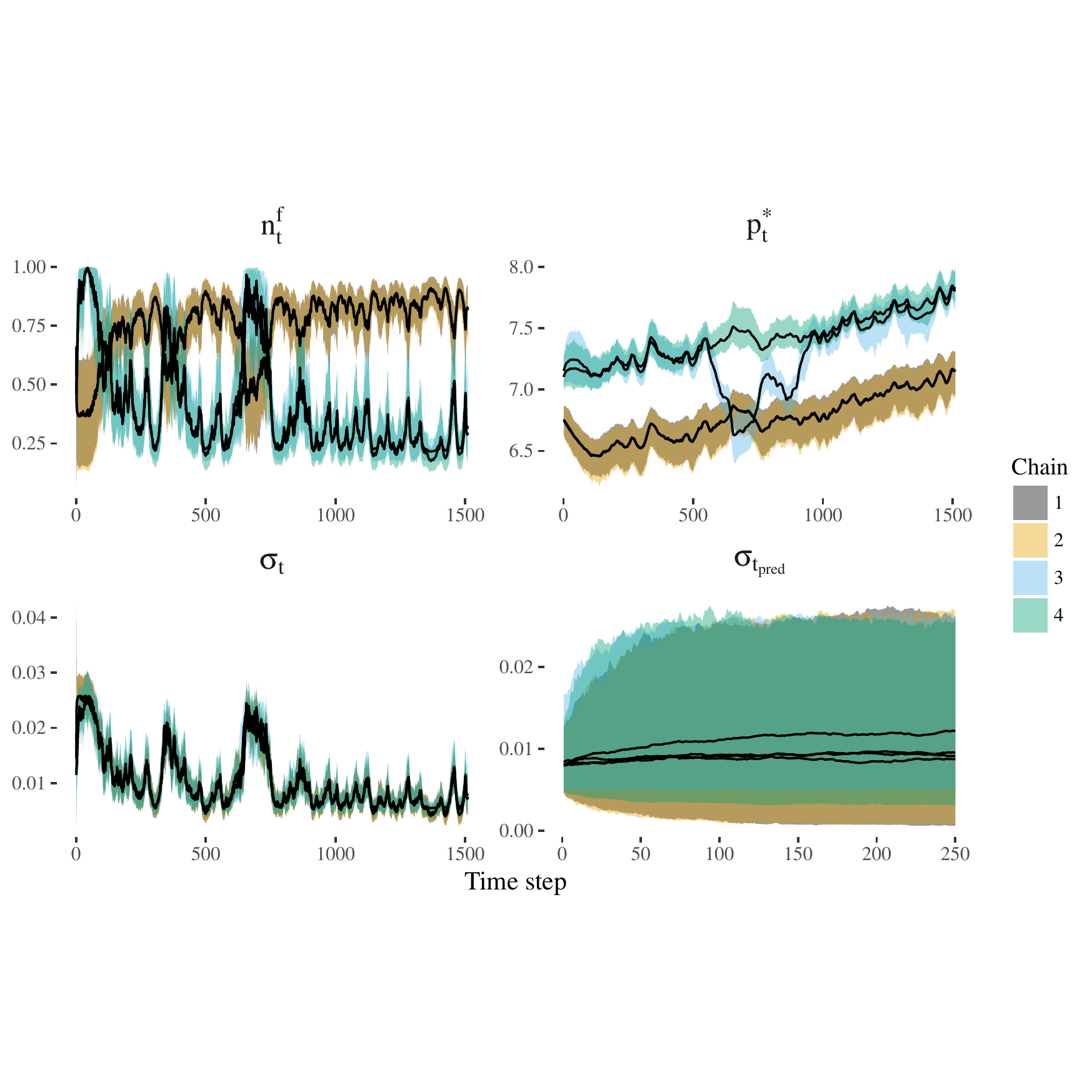}
  \caption{Trace plot (upper panel) and recovered internal states
    $n^f_t, p^*_t$ and $\sigma_t$ (lower panel) for the FW model when fitted on
    the S\&P 500. Note that the posterior appears to be
    multi-modal. Interestingly, volatile market phases either coincide
    with low or high fractions of fundamental traders with almost
    identical volatility dynamics $\sigma_t$.}
  \label{fig:FWfitSP500_multi}
\end{figure}

With simulated parameters, high
volatility arises if the fraction of chartists increases as these are
assumed to have more volatile demand, i.e.~$\sigma_c > \sigma_f$.
Looking at the trace plots for the model parameters when estimated on
S\&P 500 returns from Jan. 2009 to Dec. 2014 reveals a surprise. The
upper panel of \fig{FWfitSP500_multi} shows that chains starting from
different initial conditions have not converged to the same
distribution. This suggests a multi-modal posterior where each chain
samples from a well defined, yet different, mode of the
distribution. Furthermore, in at least one of the modes we find that
$\sigma_c < \sigma_f$!

The lower panel of \fig{FWfitSP500_multi} shows the corresponding time
series of the models state variables, i.e.~$n^f_t, p^*_t$ and
$\sigma_t$. Interestingly, the model offers two very different
explanations for the observed volatility dynamics. In the first
scenario, the number of chartists is usually low and rises sharply in
volatile market phases (as intended by the model). In
contrast, in the second scenario the number of chartists is usually
high and drops in volatile market phases. Volatility is then driven by
the high demand uncertainty of fundamental traders. Interestingly,
both scenarios lead to very similar estimates and predictions for the
volatility $\sigma_t$. The model accomplishes this by assuming a very
different trajectory for the unobserved Brownian motion of the
fundamental price leading in turn to very different mispricings and
demands of the fundamental traders. Thus, the seemingly innocuous
assumption that the fundamental price follows a Brownian motion
apparently introduces a symmetry into the model. This not only makes
the unobserved latent states of the model unidentifiable, but also reveals
that our understanding of agent-based models and their generated time
series dynamics is far from complete. Further work is certainly needed
in order characterize and ideally remove this
unidentifiability\footnote{At least if an interpretation of the
  latent state dynamics is desired. Identifiability is of no concern
  when deriving predictions.}.
\section{Conclusion}

We have demonstrated, using two different econophysics models as
examples, that agent-based models can readily be fitted with current
machine learning tools. In particular, we have implemented several models
in {\em Stan}, a modern probabilistic programming language for
Bayesian modeling. Furthermore, we showed that HMC sampling appears to
be well suited to explore the posterior distributions arising in these
agent-based models. HMC not only mixes rapidly, but also reveals clear
multi-modality in the case of the FW model.

While HMC is restricted to continuous state spaces\footnote{This
  apparent restriction can be overcome by marginalizing over discrete
  state, i.e.~summing them out. \cite{StanDoc} illustrates this
  approach on several examples including Gaussian mixture models and
  hidden Markov models.}, we have shown that different models are
either already in this form (FW model) or can be approximated
accordingly (VS model). In this regard, our work complements the study
by Lux \cite{LuxSMC} who has specifically focused on modeling
individual agents employing SMC methods to track the resulting
discrete state dynamics. We believe that HMC has several advantages.
In particular, it is highly efficient and able to handle models with
many parameters such as stochastic time varying state variables,
e.g.~the fundamental price in the FW model. Furthermore, it provides
additional diagnostics, e.g.~based on the numerical quality of the
integrated Hamiltonian dynamics, to access convergence and sample
quality.

Here, we applied HMC to estimate agent-based models by conditioning
them on several years of S\&P 500 returns. The resulting posterior
distribution allows to access latent state variables, e.g.~time
varying volatility, and derive model predictions without further
approximations. Also the precision of all estimated parameters is
readily available.  Finally, comparing model fits by their LOO
predictive likelihood, we found that the VS model appears severely
misspecified whereas the FW model is on par with purely statistical
econometric models outperforming a standard GARCH model. We are
optimistic that, given the comparable ease with which all models could
be implemented in {\em Stan}, more such attempts will be undertaken
and plan to investigate further models together with more detailed
comparisons in the future.

\section*{Acknowledgments}
NB thanks Dr.~h.~c.\ Maucher for funding his position.

\bibliographystyle{plain}
\bibliography{econostan}

\appendix
\section{Stan code}
\subsection{GARCH model}
\label{app:GARCH}
\begin{lstlisting}[style=custom,firstline=3,caption=The code for the GARCH model follows the Stan manual
\cite{StanDoc}. The main change being that we take log prices as
observed data instead of returns. The corresponding returns are then
computed as transformations of the input data. Furthermore{,} we
simulate the model forward to generate predictions conditional on the
posterior parameters. This is done in the generated quantities block
which is automatically run once after each sampling step.]
data {
  int<lower=0> N;
  vector[N] p; // log prices
  int<lower=0> N_pred;
}
transformed data {
  vector[N-1] ret = p[2:N] - p[1:(N-1)];
  real ret_sd = sqrt(variance(ret));
}
parameters {
  real<lower=0> sigma1;
  real mu;
  real<lower=0> alpha0;
  real<lower=0,upper=1> alpha1;
  real<lower=0,upper=(1-alpha1)> beta1;
}
transformed parameters {
  real<lower=0> sigma[N-1];

  sigma[1] = sigma1;
  for (i in 2:(N-1))
    sigma[i] = sqrt(alpha0
		    + alpha1 * square(ret[i-1] - mu)
		    + beta1 * square(sigma[i-1]));
}
model {
  // mu ~ normal(mean(ret), ret_sd);
  // sigma1 ~ student_t(5, 0, ret_sd);
  ret ~ normal(mu, sigma);
}
generated quantities {
  vector[N-1] log_lik;
  vector[N_pred] p_pred;
  vector[N_pred] sigma_pred;

  for (i in 1:(N-1))
    log_lik[i] = normal_lpdf(ret[i] | mu, sigma[i]);

  {
    real ret_pred;
    sigma_pred[1] = sqrt(alpha0
			 + alpha1 * square(ret[N-1] - mu)
			 + beta1 * square(sigma[N-1]));
    ret_pred = normal_rng(mu, sigma_pred[1]);
    p_pred[1] = p[N] + ret_pred;
    for (i in 2:N_pred) {
      sigma_pred[i] = sqrt(alpha0
			   + alpha1 * square(ret_pred - mu)
			   + beta1 * square(sigma_pred[i-1]));
      ret_pred = normal_rng(mu, sigma_pred[i]);
      p_pred[i] = p_pred[i-1] + ret_pred;
    }
  }
}

\end{lstlisting}

\subsection{VS model}
\label{app:VS}
\begin{lstlisting}[style=custom,firstline=4,caption=Stan code for the model by Vikram \& Sinha.
As for the GARCH model predictions are computed by simulating the model forward in the generated quantities block.]
data {
  int<lower=2> T;
  int<lower=0> T_pred;
  vector[T] p; // Note: actual price instead of log price
}
transformed data {
  vector[T] log_p = log(p);
  real max_ret = max(fabs(log_p[2:T] - log_p[1:(T-1)]));
  real ret_sd = sqrt(variance(log_p[2:T] - log_p[1:(T-1)]));
}
parameters {
  real<lower=0> mu;
  real<lower=0, upper=1> tau;  
  real<lower=0> sigma_max;
  real<lower=0> p_tau_0_raw;
  }
transformed parameters {
  real<lower=0> p_tau[T];
  vector<lower=0>[T] P_b; // Probability of trading P(|S(t)| = 1)
  real<lower=0> p_tau_0 = p[1] * p_tau_0_raw;

  // Compute running price average
  p_tau[1] = (1 - tau) * p[1] + tau * p_tau_0;
  for (t in 2:T)
    p_tau[t] = (1 - tau) * p[t] + tau * p_tau[t - 1];
  
  for (t in 1:T)  
    P_b[t] = exp(- mu * fabs(log(p[t] / p_tau[t]))); 
}
model {
  // mu ~ student_t(5, 25, 10);
  mu ~ gamma(3, 0.03);
  tau ~ beta(10, 0.5);
  p_tau_0_raw ~ normal(1, 0.05);
  sigma_max ~ normal(max_ret, max_ret / 4);
  
  log_p[2:T] ~ normal(log_p[1:(T-1)],
		      sigma_max * sqrt(2 * P_b[1:(T-1)]));
}
generated quantities {  
  vector[T-1] log_lik;
  real<lower=0> sigma[T];
  real p_pred[T_pred];
  real p_tau_pred[T_pred];
  real P_b_pred[T_pred];
  real sigma_pred[T_pred];

  for (t in 1:T)
    sigma[t] = sigma_max * sqrt(2 * P_b[t]);
  
  for (t in 1:(T-1))
    log_lik[t] = normal_lpdf(log_p[t+1] | log_p[t], sigma[t]);

  {
    real ret;
    
    ret = normal_rng(0, sigma[T]);
    p_pred[1] = exp(ret) * p[T];
    
    p_tau_pred[1] = (1 - tau) * p_pred[1] + tau * p_tau[T];
    P_b_pred[1] = exp(- mu * fabs(log(p_pred[1] / p_tau_pred[1])));  
    sigma_pred[1] = sigma_max * sqrt(2 * P_b_pred[1]);
  
    for (t in 2:T_pred) {  
      ret = normal_rng(0, sigma_pred[t-1]);
      p_pred[t] = exp(ret) * p_pred[t-1];
      p_tau_pred[t] = (1 - tau) * p_pred[t] + tau * p_tau_pred[t-1];
      P_b_pred[t] = exp(- mu * fabs(log(p_pred[t] / p_tau_pred[t])));
      sigma_pred[t] = sigma_max * sqrt(2 * P_b_pred[t]);  
    }
  }
}

\end{lstlisting}

\subsection{FW model}
\label{app:FW}
\begin{lstlisting}[style=custom,firstline=3,caption=Stan code for Franke
\& Westerhoff model. Agent dynamics follows the DCA-HPM specification and
the fundamental log price $p^*_t$ is modeled as a Brownian
motion. Note the use of the non-centered parameterization for this random walk.]
data {
  int<lower=1> N;
  vector[N] p;   // log prices
  int<lower=0> N_pred;
}
transformed data {
  // mu and beta fixed ... redundant anyways
  real mu = 0.01;
  real beta = 1.0;

  real ret_sd = sqrt(variance(p[2:N] - p[1:(N-1)]));
}
parameters {
  real<lower=0> phi;
  real<lower=0> xi;
  real alpha_0;
  real<lower=0> alpha_n;
  real<lower=0> alpha_p;
  real<lower=0> sigma_f;
  real<lower=0> sigma_c;
  real<lower=0, upper=1> n_f_1;
  // p_star random walk in non-centered parameterization
  vector[N] epsilon_star;
  real<lower=0> sigma_p_star;
}
transformed parameters {
  vector[N] n_f;
  vector[N] demand;
  vector[N] sigma;
  vector[N] p_star;

  p_star[1] = p[1] + 5.0 * sigma_p_star * epsilon_star[1];
  for (i in 2:N)
    p_star[i] = p_star[i-1] + sigma_p_star * epsilon_star[i];
  
  n_f[1] = n_f_1;
  demand[1] = 0;
  sigma[1] = mu * sqrt( square(n_f[1] * sigma_f)
			+ square((1 - n_f[1]) * sigma_c));
  for (i in 2:N) {
    // equation (HPM)
    real a = alpha_n * (n_f[i-1] - (1 - n_f[i-1]))
      + alpha_0
      + alpha_p * square(p[i-1] - p_star[i-1]);
    // equation (DCA)
    n_f[i] = inv_logit(beta * a);
    demand[i] = mu * (n_f[i] * phi * (p_star[i] - p[i])
		      + (1 - n_f[i]) * xi * (p[i] - p[i-1]));
    // structured stochastic volatility
    sigma[i] = mu * sqrt( square(n_f[i] * sigma_f)
			  + square((1 - n_f[i]) * sigma_c));
  }
}
model {
  phi ~ student_t(5, 0, 1);
  xi ~ student_t(5, 0, 1);
  alpha_0 ~ student_t(5, 0, 1);
  alpha_n ~ student_t(5, 0, 1);
  alpha_p ~ student_t(5, 0, 1);
  sigma_f ~ normal(0, ret_sd / mu);
  sigma_c ~ normal(0, 2.0 * ret_sd / mu);
  epsilon_star ~ normal(0, 1);
  sigma_p_star ~ normal(0, ret_sd / 2.0);
  
  // Price likelihood
  p[2:N] ~ normal(p[1:(N-1)] + demand[1:(N-1)], sigma[1:(N-1)]);
}
generated quantities {
  vector[N-1] log_lik;
  vector[N_pred] p_pred;
  vector[N_pred] n_f_pred;
  vector[N_pred] demand_pred;
  vector[N_pred] sigma_pred;
  vector[N_pred] p_star_pred;

  for (i in 1:(N-1))
    log_lik[i] = normal_lpdf(p[i+1] | p[i] + demand[i], sigma[i]);

  if (N_pred > 0) {
    real a;

    // predict fundamental price walk
    p_star_pred[1] = normal_rng(p_star[N], sigma_p_star);
    for (i in 2:N_pred)
      p_star_pred[i] = normal_rng(p_star_pred[i-1], sigma_p_star);

    p_pred[1] = normal_rng(p[N] + demand[N], sigma[N]);
    a = alpha_n * (n_f[N] - (1 - n_f[N])) + alpha_0 + alpha_p * square(p[N] - p_star[N]);
    n_f_pred[1] = inv_logit(beta * a);
    demand_pred[1] = mu * (n_f_pred[1] * phi * (p_star_pred[1] - p_pred[1])
			   + (1 - n_f_pred[1]) * xi * (p_pred[1] - p[N]));
    sigma_pred[1] = mu * sqrt( square(n_f_pred[1] * sigma_f)
			       + square((1 - n_f_pred[1]) * sigma_c));
    for (i in 2:N_pred) {
      p_pred[i] = normal_rng(p_pred[i-1] + demand_pred[i-1], sigma_pred[i-1]);
      a = alpha_n * (n_f_pred[i-1] - (1 - n_f_pred[i-1])) + alpha_0 + alpha_p * square(p_pred[i-1] - p_star_pred[i-1]);
      n_f_pred[i] = inv_logit(beta * a);
      demand_pred[i] = mu * (n_f_pred[i] * phi * (p_star_pred[i] - p_pred[i])
			   + (1 - n_f_pred[i]) * xi * (p_pred[i] - p_pred[i-1]));
      sigma_pred[i] = mu * sqrt( square(n_f_pred[i] * sigma_f)
				 + square((1 - n_f_pred[i]) * sigma_c));
    }
  }
}
\end{lstlisting}

\end{document}